\documentclass[alpha-refs]{nbdt-article}

\usepackage{graphicx}
\usepackage{natbib}
\usepackage{multibib}
\usepackage{ragged2e}

\usepackage{url} 
\usepackage{float}
\usepackage[table]{xcolor}
\definecolor{dred}{rgb}{0.6,0,0}
\definecolor{dpurple}{HTML}{A020F0}
\definecolor{dblue}{rgb}{0,0,0.6}
\definecolor{hlcolor}{rgb}{1,1,0.8}

\usepackage{caption}
\usepackage{subcaption}
\usepackage{amsmath,amssymb,amsfonts,bm}
\usepackage{multicol}
\usepackage{lineno}
\usepackage{authblk}
\usepackage{hyperref} 
\usepackage{cleveref}
\usepackage{geometry}
\renewcommand\b\bm

\Crefname{equation}{Equation}{Equations}
\Crefname{figure}{Figure}{Figures}
\creflabelformat{equation}{#2#1#3}
\crefrangelabelformat{equation}{#3#1#4-#5#2#6}

\usepackage{titlesec}
\titlespacing\section{0pt}{10pt plus 3pt minus 2pt}{2pt plus 2pt minus 2pt}
\titlespacing\subsection{0pt}{6pt plus 2pt minus 2pt}{1pt plus 2pt minus 2pt}
\titlespacing\subsubsection{0pt}{4pt plus 1pt minus 1pt}{0pt plus 2pt minus 2pt}

\title{An introduction to reinforcement learning for neuroscience}
\author[1,2]{Kristopher T. Jensen}
\affil[1]{Sainsbury Wellcome Centre, University College London}
\affil[2]{Computational and Biological Learning Lab, University of Cambridge}
\corraddress{Kristopher Jensen, Sainsbury Wellcome Centre, University College London, London W1T 4JG, UK}
\corremail{kris.torp.jensen@gmail.com}
\runningauthor{Jensen}

\begin{document}

\maketitle

\begin{abstract}
\noindent
Reinforcement learning (RL) has a rich history in neuroscience, from early work on dopamine as a reward prediction error signal \citep{schultz1997neural} to recent work proposing that the brain could implement a form of `distributional reinforcement learning' popularized in machine learning \citep{dabney2020distributional}.
There has been a close link between theoretical advances in reinforcement learning and neuroscience experiments throughout this literature, and the theories describing the experimental data have therefore become increasingly complex.
Here, we provide an introduction and mathematical background to many of the methods that have been used in systems neroscience.
We start with an overview of the RL problem and classical temporal difference algorithms, followed by a discussion of `model-free', `model-based', and intermediate RL algorithms.
We then introduce deep reinforcement learning and discuss how this framework has led to new insights in neuroscience.
This includes a particular focus on meta-reinforcement learning \citep{wang2018prefrontal} and distributional RL \citep{dabney2020distributional}.
Finally, we discuss potential shortcomings of the RL formalism for neuroscience and highlight open questions in the field.
Code that implements the methods discussed and generates the figures is also \href[]{https://colab.research.google.com/drive/1ZC4lR8kTO48yySDZtcOEdMKd3NqY_ly1?usp=sharing}{provided}.
\end{abstract}

\newpage

$ $\vspace{-2.4em}

\section{Introduction}
\label{sec:intro}

Humans and other animals learn from their experiences.
Sometimes, this takes the form of explicit demonstration, as is often the case during formal education.
However, we also often have to learn from trial and error together with feedback received from the world around us -- sometimes implicit and sometimes explicit.
This is well illustrated by the classical case study of Pavlov's dogs, who learned to associate a so-called `conditioned stimulus' (CS; e.g. the ringing of a bell) with the availability of food shortly after (the `unconditioned stimulus'; US).
Following a brief period of learning, the dogs would start to salivate in response to the CS in advance of any food actually being served.
This suggests that the dogs had learned to associate the CS with the availability of `reward' in the form of food, and that they produced an appropriate physiological response to take advantage of this food availability.
Importantly, this occurred without any explicit instruction or description of the sequence of events preceding food being served.
Instead, the dogs learned from experience with their environment and the presence of a salient, rewarding stimulus.

Such passive stimulus-response predictions are also called `Pavlovian learning' and have been commonly used in neuroscience to study learning from external rewards \citep{niv2009reinforcement}.
This forms a specific instantiation of the concept of `reinforcement learning', which is a general term for settings where an agent learns from reward signals in the environment rather than explicit demonstration, as is the case in `supervised learning'.
Importantly, the past decades have shown that principles of reinforcement learning can be used to explain not just behaviour, but also neural activity in biological circuits \citep{niv2009reinforcement,botvinick2020deep}.
An explicit neural basis of RL was initially demonstrated in foundational work by \citet{schultz1997neural}, which showed that the firing rates of dopaminergic neurons in the Ventral Tegmental Area (VTA) reflected the difference between expected and actual `value' when animals received a juice reward following a CS consisting of a lever-press in response to a small light turning on.
This provided a potential neural substrate of the classical `temporal difference' learning algorithm \citep{schultz1997neural,sutton1988learning}, which has since been expanded to a wealth of evidence for reinforcement learning in neural dynamics \citep{niv2009reinforcement, dabney2020distributional,watabe2017neural}.
However, these classical algorithms are generally restricted to simple problem settings, while humans and other animals are capable of solving complex high-dimensional problems involving extended planning and fine motor control.
The field of `deep reinforcement learning' has recently emerged to tackle such problems in a machine learning setting, which has led to impressive results across a range of tasks \citep{mnih2013playing, schrittwieser2020mastering, wurman2022outracing, vinyals2019grandmaster}.
Intriguingly, recent research has demonstrated that these deep RL algorithms also have parallels in both behaviour and neural dynamics \citep{botvinick2020deep, wang2018prefrontal, dabney2020distributional, jensen2023recurrent,aldarondo2024virtual}, suggesting that neuroscience can continue to learn from advances in reinforcement learning.

In this review, we provide an overview of the reinforcement learning problem and popular algorithms, with a particular focus on parallels and uses of these algorithms in neuroscience.
This overview starts from classical tabular TD learning and Q-learning algorithms, which have guided neuroscience research for decades.
We then consider the important distinction between model-based and model-free reinforcement learning, as well as methods that fall somewhere in the gray area between these extremes, and discuss their neural correlates.
Finally, we generalize the tabular methods to the non-linear function approximation setting and the resulting deep RL methods, which have revolutionized machine learning in recent years.
We do this with a focus on methods that have had a strong influence on neuroscience to give the reader a better idea of the mathematical and computational background of recent neuroscientific findings.
These include the `meta-reinforcement learning' model of PFC by \citet{wang2018prefrontal} and the `distributional reinforcement learning' model of VTA dopaminergic neurons by \citet{dabney2020distributional} in particular.
We hope this review will be useful both for those who are interested in the theory underlying reinforcement learning in neuroscience and for those who want an overview of how the neuroscience literature builds on principles from reinforcement learning theory.
Throughout the paper, the focus will be on an intuitive understanding of the relevant RL methods, and explicit derivations are included only where we consider them conducive to such understanding.
We refer to \citet{sutton2018reinforcement} for a more in-depth treatment of the underlying theory.

\section{Problem setting}\label{sec:problem_setting}

Here we provide a short introduction to the reinforcement learning problem in a discrete state and action space with a finite time horizon -- a common setting for neuroscience experiments consisting of repeated trials or episodes in a controlled environment.
In this setting, the environment consists of states from a discrete set $s \in \mathcal{S} = {\{ s_1, s_2, \ldots \}}_1^{|\mathcal{S}|}$, and the agent can take actions $a \in \mathcal{A} = {\{ a_1, a_2, \ldots \}}_1^{|\mathcal{A}|}$.
The environment is characterized by transition and reward probabilities $p(s_{t+1}, r_t | s_t, a_t)$, where $r_t$ is the reward at time $t$.
We will use $r(s, a)$ to denote either the reward when it depends deterministically on the state and action, or its expectation otherwise.
We will further make the \emph{Markov assumption} that the next state and reward only depend on the current state and action, $p(s_{t+1}, r_t | s_t, a_t, s_{t-1}, a_{t-1}, \ldots, s_0, a_0) = p(s_{t+1}, r_t | s_t, a_t)$.

\begin{figure*}[!t]
    \centering
    \vspace*{-0.5em}
    \includegraphics[width=0.85\textwidth]{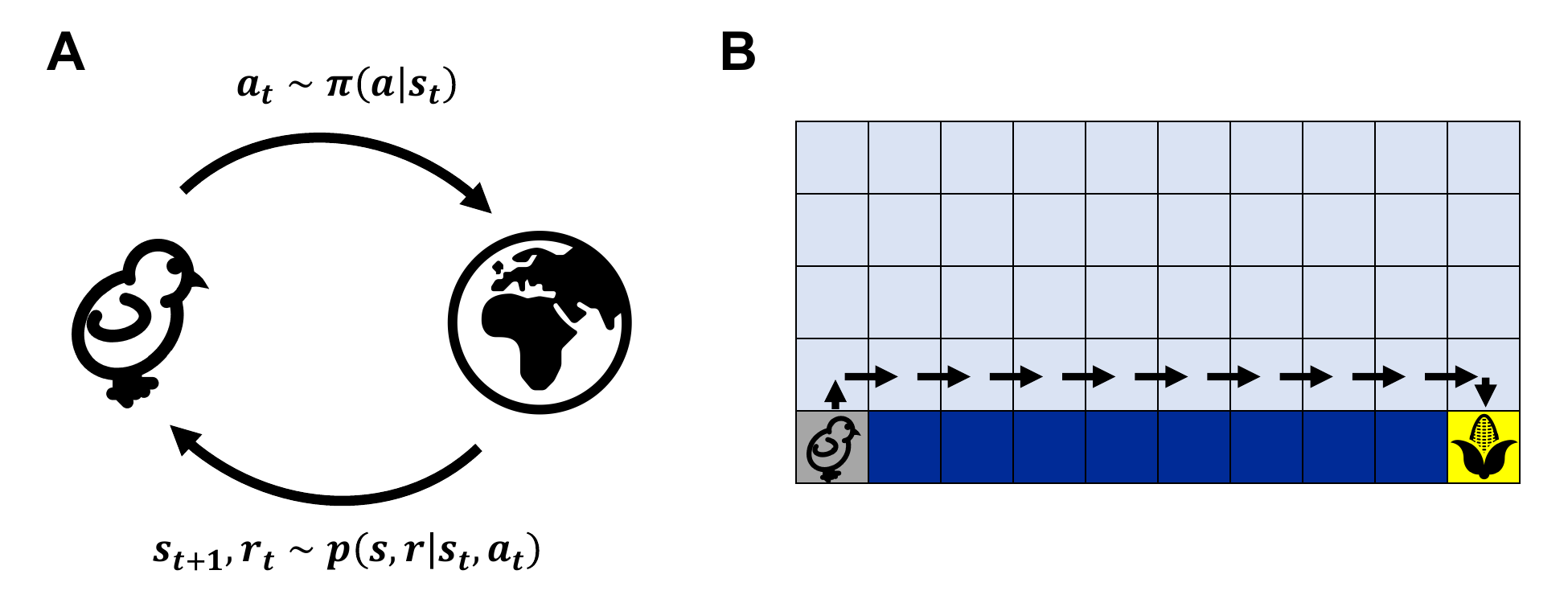}
    \vspace*{-1.0em}
    \caption[RL schematics]{\label{fig:schematics} \justifying
        {\bfseries The reinforcement learning problem and cliffworld environment.}
        {\bfseries (A)}~An agent (here the bird) interacts with the world to maximize reward.
        This involves a balance between exploring potentially interesting new states (e.g. searching for food in a new field) while also exploiting states known to yield high reward (e.g. the field that had many worms yesterday).
        At a given point in time, the bird is in some state $s_t$ from which it can take an action $a_t$, with the probability of different actions determined by the `policy' $\pi(a|s_t)$, which is controlled by the agent.
        $a_t$ then leads to a change in the environment according to the non-controllable environment dynamics $s_{t+1}, r_t \sim p(s, r | s_t, a_t)$.
        Here, $r_t$ is the empirical `reward' received by the agent, and its objective is to collect as much cumulative reward as possible.
        Often, reinforcement learning problems are divided into `episodes', with the agent learning over the course of multiple repeated exposures to the environment.
        This could for example consist of the bird learning over multiple days which fields are likely to be rich in food, while minimizing the distance travelled and exposure to predators.
        {\bfseries (B)}~The `cliffworld' environment, which will be used to demonstrate the performance and behaviour of a range of reinforcement learning algorithms in this work.
        The agent starts in the lower left corner (location [0, 0]), and the episode finishes when it encounters either the `cliff' (dark blue) or the goal (yellow; location [9,0]).
        If the agent walks off the cliff, it receives a reward of -100.
        If it finds the goal, it receives a reward of +50.
        In any other state, it receives a reward of -1.
        Such negative rewards for `neutral' actions are commonly used to encourage the agent to achieve its goal as fast as possible.
        The arrows indicate the `optimal' policy, which takes the agent to the goal via the shortest possible route that avoids the cliff.
        }
    \vspace*{-0.5em}
\end{figure*}

We can now define a \emph{trajectory} $\tau = {\{ s_t, a_t, r_t \}}_{t = 0}^T$.
The probability of a trajectory occuring is
\begin{linenomath*}
\begin{equation}
    p_\pi(\tau) = p(s_0) \prod_{t = 0}^T p(s_{t+1}, r_t | s_t, a_t) \pi(a_t|s_t).
\end{equation}
\end{linenomath*}
$p_\pi(\tau)$ depends on the \emph{policy} of the agent, $\pi(a|s)$, which specifies the probability of taking action $a$ in state $s$ (\Cref{fig:schematics}A).
The objective is to learn a policy that maximizes the expected total discounted reward
\begin{linenomath*}
\begin{align}
    \label{eq:RL_objective}
    J(\pi) &= \mathbb{E}_{\tau \sim p_\pi(\tau)} \left [ R_\tau \right ] = \mathbb{E}_{\tau \sim p_\pi(\tau)} \left [ \sum_{t=0}^T \gamma^t r_t | \tau \right ],
\end{align}
\end{linenomath*}
where $R_\tau := \sum_{t=0}^T \gamma^t r_t | \tau$, and we have written $J(\pi)$ since the policy uniquely specifies $J$ in a stationary environment.
In \Cref{eq:RL_objective}, $\gamma$ is a `discount factor', which stipulates that we should care more about immediate rewards than rewards far in the future.
We can provide three interpretations for this discount factor.
One is that agents intrinsically care more about immediate reward than distant reward.
A second is that there is a fixed non-zero probability $(1-\gamma)$ of the current `episode' or environment terminating or changing at each timestep, in which case we should weight putative future reward by the probability that we are still engaged in the task at that time.
The third view is that $\gamma$ simply provides a tool for reducing the variance of our learning methods, especially in temporally extended tasks.
This third view is most compatible with the fact that \emph{evaluation} of RL agents after training is generally done without discounting.

Since $J(\pi$) depends on the policy of the agent, it is possible to search in the space of policies for one that maximizes $J$, which is the topic of reinforcement learning.
It is often assumed that the experience $\{ \tau \}$ is generated by the agent acting according to its policy, and the resulting experience is then used to update the policy in a way that increases $J(\pi)$.
However, `off-policy' and `offline' reinforcement learning methods also exist, where the agent learns on the basis of experience generated by a policy different from $\pi$ (\citealp{levine2020offline}; \Cref{sec:off-policy}).
This can be either an `old' version of the agent itself, when it acted according to a different policy, or data generated by an entirely different agent.
Off-policy learning is important for biological organisms, where learning can happen `offline' during sleep after initial data collection during wake, or from observing other individuals (also the topic of `imitation learning'; \Cref{sec:imitation}).

\section{Temporal difference learning}\label{sec:temporal_difference}

A simple way to maximize reward in an environment is to learn the `value' of different states, and then move towards states with high value.
The potential importance of such an algorithm for neuroscience is evident from the value-seeking behaviour of many organisms, and the widespread findings of neural `codes' for value across the brain \citep{schultz1992neuronal, padoa2006neurons, rushworth2011frontal}.
This leaves the question of how such value codes can be learned in a biologically plausible setting.

\begin{figure*}[!t]
    \centering
    \vspace*{-0.5em}
    \includegraphics[width=0.95\textwidth]{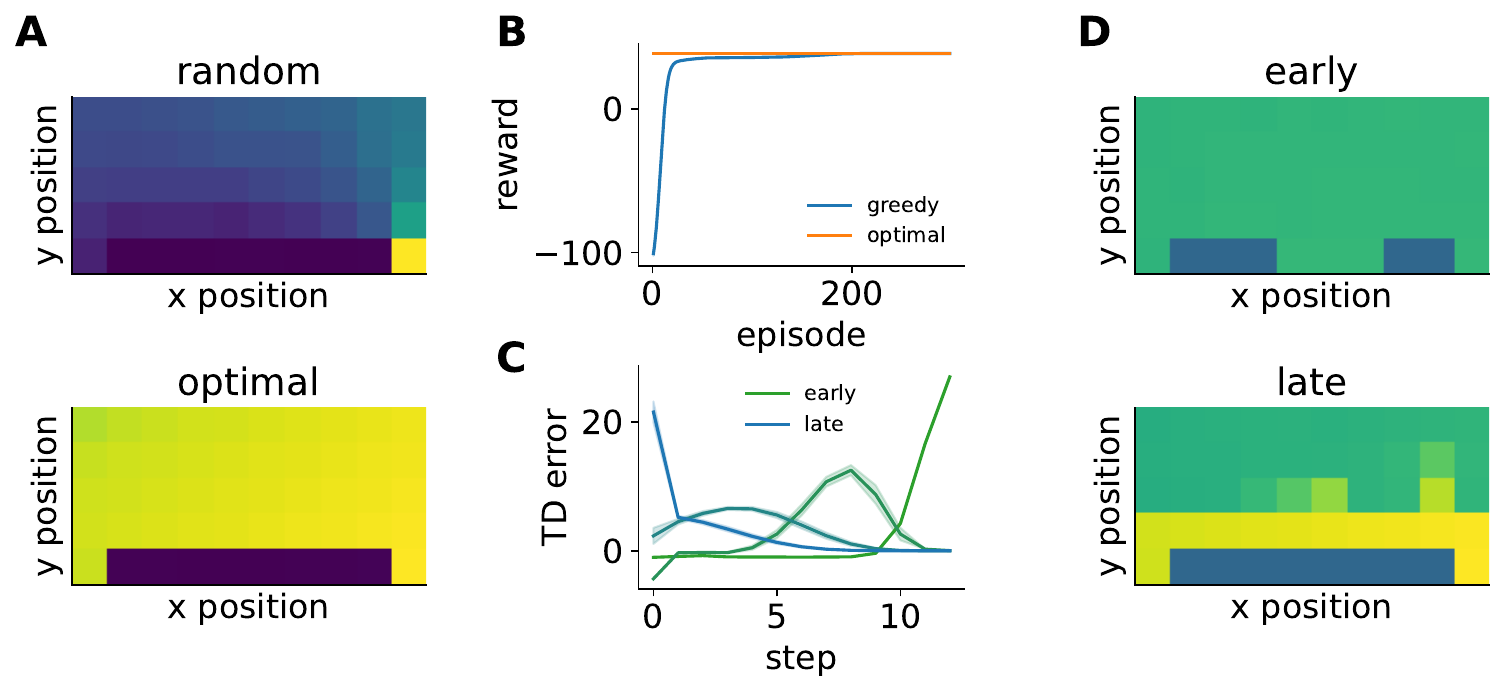}
    \vspace*{-1.0em}
    \caption[TD]{\label{fig:TD} \justifying
        {\bfseries Temporal difference learning.}
        {\bfseries (A)}~Value functions aquired through temporal difference learning (\Cref{eq:TD-learning}) while acting according to either a random (top) or an optimal (bottom) policy.
        These simulations were performed with a random start state in the cliffworld environment to ensure full coverage of the space.
        Dark blue indicates negative expected reward (-100) and yellow indicates positive expected reward (+50).
        These simulations used a learning rate of $\alpha = 0.05$ and no temporal discounting ($\gamma = 1$).
        Under the random policy, states near the cliff have low value even if they are close to the goal, since the agent often falls off the cliff from there.
        Under the optimal policy, all states have high expected reward, since the agent always reaches the goal.
        States nearer the goal have slightly higher value than those further away.
        {\bfseries (B)}~Empirical reward as a function of episode number for a TD-learning agent that acts according to \Cref{eq:value_action_selection} while updating its value estimates according to \Cref{eq:TD-learning}.
        For this agent, action selection assumes access to a `one-step' world model in order to evaluate the consequence of each putative action.
        The agent gradually converges to an optimal policy.
        Parameters for the agent are as in (A), except that the start state is always the lower left corner.
        {\bfseries (C)}~TD error (\Cref{eq:TD-learning}) as a function of the step number along the optimal path for the agent in (B) at different stages of learning (green to blue).
        This TD signal gradually propagates backwards from the reward to preceding states, mirroring biological recordings of dopamine activity \citep{schultz1997neural}.
        {\bfseries (D)}~Value function learned by a greedy TD agent as in (B), plotted either early (top) or late (bottom) in training.
        Early in training, the agent has learned that the cliff is bad but doesn't know where the goal is or how to get there.
        Late in training, the agent has learned a value function that locally resembles the optimal value function from (A), while it has not learned the value of distant states that are rarely or never visited from the start state.
        This is a potential shortcoming of `greedy' agents that can easily converge to a sub-optimal local maximum in more complicated environments.
        For this analysis, we used a high learning rate of $\alpha = 0.5$ to make the early TD updates larger and therefore more visible.
        }
    \vspace*{-0.5em}
\end{figure*}

One answer to this question takes the form of the classical `temporal difference learning' algorithm \citep{sutton1988learning,sutton2018reinforcement}.
This involves defining a \emph{value function} for a given state $s$ and policy $\pi$, which quantifies the expected future reward when following $\pi$ starting from $s$:
\begin{linenomath*}
\begin{equation}
    \label{eq:V-values}
    V^{\pi}(s) = \mathbb{E}_{\tau \sim p_\pi(\tau)} \left [ \sum_{t' \geq t} \gamma^{t' - t} r_{t'} | s_t = s \right ].
\end{equation}
\end{linenomath*}
Here, $\mathbb{E}_{\tau \sim p_\pi(\tau)} [ \cdot ]$ indicates an expectation taken over trajectories $\tau$ resulting from the agent following the policy $\pi$.
For the true value function, we can expand this as a self-consistency equation
\begin{linenomath*}
\begin{align}
    V^{\pi}(s) &= r_\pi(s) + \sum_{s'} p_{\pi}(s_{t+1} = s' | s_t = s) \mathbb{E}_{\tau \sim p_\pi(\tau)} \left [ \sum_{t' = t+1} \gamma^{t' - t} r_{t'} | s_{t+1} = s' \right ] \\
    &=  r_\pi(s) + \gamma \sum_{s'} p_{\pi}(s_{t+1} = s' | s_t = s) V^{\pi}(s'),
    \label{eq:value_expansion}
\end{align}
\end{linenomath*}
where $p_{\pi}(s_{t+1} = s' | s_t = s) = \sum_a \pi(a|s) p(s_{t+1} = s' | s_t = s, a_t = a)$ is the probability of transitioning from $s$ to $s'$ under $\pi$, and $r_\pi(s) = \mathbb{E}_{a \sim \pi} [ r(s, a) ]$ is the expected reward in state $s$, averaged over actions.
Importantly, \Cref{eq:value_expansion} would not hold if $V^{\pi}(s)$ was not the true value function \citep{sutton2018reinforcement}.
When learning an approximate value function $V(s)$, we can therefore use this bootstrapped self-consistency relation as an objective function:
\begin{linenomath*}
\begin{equation}
    \mathcal{L} \propto {\left ( V(s) - \left ( r_\pi(s) + \gamma \mathbb{E}_{p_\pi(s'|s)} \left [ V(s') \right ] \right ) \right )}^2,
\end{equation}
\end{linenomath*}
Gradient descent w.r.t $V(s)$ gives us an update rule
\begin{linenomath*}
\begin{align}
    \Delta V(s) &\propto - \frac{\partial \mathcal{L}}{\partial V(s)}\\
    \label{eq:TD-learning_exp}
    &\propto - V(s) + r_\pi(s) + \gamma \mathbb{E}_{p_\pi(s'|s)} \left [ V(s') \right ]\\
    \label{eq:TD-learning}
    &\approx - V(s_t) + r_t + \gamma V(s_{t+1}).
\end{align}
\end{linenomath*}
The last line approximates the expected update with a single sample corresponding to the states and reward actually experienced.
As more experience is collected and many small gradient steps are taken according \Cref{eq:TD-learning}, these single-sample estimates average out to the expectation in \Cref{eq:TD-learning_exp} (\Cref{fig:TD}A).
Variants of this algorithm can also learn about multiple past states at once using the notion of \emph{eligibility traces} \citep{sutton2018reinforcement}.
However, \Cref{eq:TD-learning} is the canonical temporal difference learning rule \citep{sutton1988learning}, and it leads to learning dynamics where the temporal difference error $\delta_t := - V(s_t) + r_t + \gamma V(s_{t+1})$ gradually propagates from the rewarding state to prior states that predict the upcoming reward (\Cref{fig:TD}C).

This gradual propagation of prediction errors from the reward state to its predecessors has been of great interest in neuroscience.
In particular, classical work by \citet{schultz1997neural} demonstrated a similar pattern of neural activity in dopaminergic VTA neurons, which formed the foundation of a now well-established theory that dopamine provides a biological reward prediction error signal that drives learning \citep{niv2009reinforcement,watabe2017neural}.
At a behavioural level, this is supported by experiments showing that artificial stimulation of dopamine neurons can be a strong driver of learning \citep{olds1954positive, tsai2009phasic, steinberg2013causal}.
The simple narrative of dopamine as a reward prediction error has also been challenged in recent years \citep{coddington2019learning, howe2013prolonged, horvitz2000mesolimbocortical}, which has led to theories of dopamine as a more general prediction error for both value but also other quantities like salience and movement \citep{kakade2002dopamine, gershman2024explaining}.

Much classical work in neuroscience has focused on value learning in Pavlovian conditioning tasks, but animals in natural environments also have to take actions on the basis of this information.
However, after learning a value function, it can be used for optimal action selection if we can estimate $r(s, a)$ and $p(s' | s, a)$.
We can then compute a `state-action value' $Q^\pi(s ,a)$, defined as the expected discounted future reward associated with taking action $a$ in state $s$ and then following policy $\pi$:
\begin{linenomath*}
\begin{equation}
    \label{eq:Q-values}
    Q^\pi(s,a) := \mathbb{E} \left [\sum_{t' = t} \gamma^{t'-t} r_{t'} | s_t = s, a_t = a \right ] = r(s, a) + \gamma \sum_{s'} p( s' | s, a) V^\pi(s').
\end{equation}
\end{linenomath*}
Approximating the true value function $V^\pi$ by the learned value function $V$ in \Cref{eq:Q-values} yields an approximate state-action value $Q$, which can be used to choose the action with the highest expected reward,
\begin{linenomath*}
    \begin{equation}
        \label{eq:value_action_selection}
        a^*(s) = \text{argmax}_{a} Q(s, a).
    \end{equation}
\end{linenomath*}
Updating the value function according to \Cref{eq:TD-learning} while acting in the environment according to \Cref{eq:value_action_selection} leads to an agent that gradually learns to take better actions as it learns a better value function (\Cref{fig:TD}B-D).
This provides a biologically plausible algorithm for reward-driven learning in agents with access to a one-step predictive model.

\section{Q-learning}
\label{sec:q_learning}

In some cases, we may not know the transition function or it could be expensive to simulate.
Additionally, it has been found that dopamine activity can reflect learning signals not just for state values but also for \emph{action} values \citep{roesch2007dopamine,morris2006midbrain}.
This suggests an alternative model of biological learning, where animals directly learn the state-action values defined in \Cref{eq:Q-values}.
This is in contrast to the algorithm in \Cref{sec:temporal_difference}, where the agent only learned the \emph{state} values, and then computed the action values at decision-time using a one-step world model.
Q-learning is the most prominent model for learning state-action values, and it has commonly been used to explain animal behaviour and neural activity \citep{niv2009reinforcement, mattar2018prioritized}.

\begin{figure*}[!t]
    \centering
    \vspace*{-0.5em}
    \includegraphics[width=0.95\textwidth]{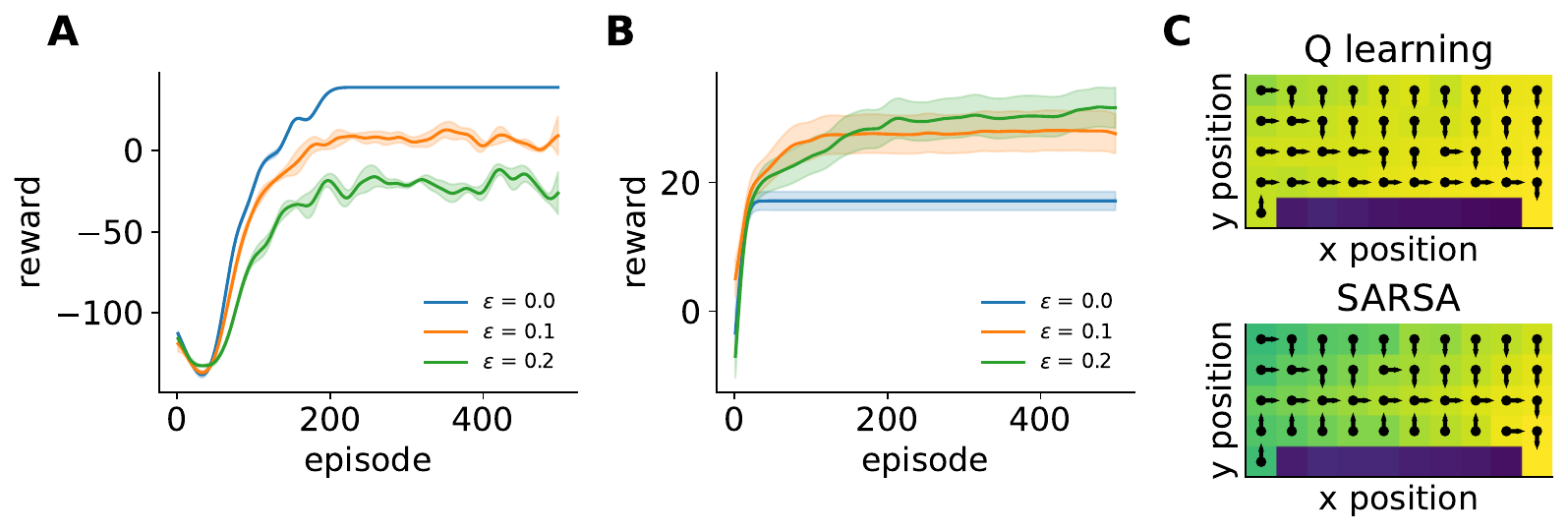}
    \vspace*{-1.0em}
    \caption[Q]{\label{fig:Q} \justifying
        {\bfseries Q-learning.}
        {\bfseries (A)}~Empirical reward as a function of episode number for Q-learners with different levels of stochasticity in their policy ($\epsilon \in \{0, 0.1, 0.2\}$; legend).
        For these simulations, we used a learning rate of $\alpha = 0.05$ for all agents and no temporal discounting ($\gamma = 1$).
        The agent with $\epsilon = 0$ converges to an optimal policy, similar to the TD agent in \Cref{fig:TD}A.
        However, convergence is in this case slower despite using the same learning rate, because the Q-learner has to learn about each action independently, while the TD agent used its one-step world model to aggregate learning across actions reaching the same state.
        In this cliffworld environment, increasing epsilon leads to worse performance since it increases the probability of falling off the cliff.
        Additionally, there is no risk of getting stuck in a local minimum since there is only one rewarding state, which decreases the value of exploration.
        Lines and shading indicate mean and standard error across 10 simulations.
        {\bfseries (B)}~As in (A), now for a non-cliffworld grid environment with two goals: one with a reward of +20 at location (0, 4), and one with a reward of +50 at location (5,0).
        In this case, having non-zero epsilon can increase the probability of discovering the `high reward' goal rather than getting stuck with a locally optimal policy of moving to the `low reward' goal.
        In these simulations, we used a learning rate of $\alpha = 1$, since this effect is less robust with lower learning rates that lead to more exploration of the environment across all agents.
        {\bfseries (C)}~Cliffworld policy learned by a Q-learning (top) or SARSA (bottom) agent with $\epsilon = 0.3$.
        Colours indicate the maximum value of any action in a state from blue (-100) to yellow (+50), and arrows indicate which action has the highest value.
        The Q-learning agent learns to move right above the cliff, because this is the optimal thing to do under the assumption that subsequent actions are also optimal.
        This is because it is an `off-policy' algorithm that does not take into account the actual policy of the agent.
        In contrast, the SARSA agent learns to move a `safe distance' away from the cliff, since it is an `on-policy' algorithm that takes into account the finite probability of the agent choosing to move off the cliff from upcoming states.
        Q-learning agents are also frequently trained using a stochastic $\epsilon$-greedy policy and then evaluated with the greedy policy corresponding to $\epsilon = 0$, or they can be trained while `annealing' $\epsilon$ from some finite value to $0$ over several episodes to allow for initial exploration.
        }
    \vspace*{-0.5em}
\end{figure*}

To learn the Q-values necessary for action selection directly, we start by expanding \Cref{eq:Q-values},
\begin{linenomath*}
\begin{equation}
    \label{eq:Q-expanded}
    Q^\pi(s,a) = r(s, a) + \gamma \sum_{s'} p(s_{t+1} = s'| s_t=s, a_t=a) \sum_{a'} \pi(a' | s') Q^\pi(s', a').
\end{equation}
\end{linenomath*}
For the greedy policy $\pi^g(a|s) := I_{a = a^*(s)}$ (where the indicator function $I_{a = b} = 1$ for $a=b$ and $0$ otherwise), this gives rise to a self-consistency expression for the state-action values:
\begin{linenomath*}
\begin{equation}
    \label{eq:Q-optimal}
    Q^{\pi^g}(s,a) = r(s, a) + \gamma \mathbb{E}_{s' \sim p(s' | s, a)} \left [ \text{max}_{a'} Q^{\pi^g}(s', a') \right ].
\end{equation}
\end{linenomath*}
Importantly, this self-consistency expression only holds when the Q-values have converged to the true expected rewards, and the associated greedy policy is therefore optimal \citep{sutton2018reinforcement}.
We can now use \Cref{eq:Q-optimal} as an objective by defining
\begin{linenomath*}
\begin{equation}
    \mathcal{L} \propto {\left (  Q(s,a) - \left (r(s, a) + \gamma \mathbb{E}_{s' \sim p(s' | s, a)} \left [ \text{max}_{a'} Q(s', a') \right ] \right ) \right )}^2,
\end{equation}
\end{linenomath*}
Gradient descent w.r.t $Q(s,a)$ gives us an update rule
\begin{linenomath*}
\begin{align}
    \Delta  Q(s,a) & \propto - Q(s,a) + r(s, a) + \gamma \mathbb{E}_{s' \sim p(s' | s, a)} \left [ \text{max}_{a'} Q(s',a') \right ] \\
    &\approx - Q(s_t,a_t) + r_t + \gamma \text{max}_{a'} Q(s_{t+1}, a').
\end{align}
\end{linenomath*}
This is the so-called Q-learning update rule (\citealp{watkins1989learning}; \Cref{fig:Q}A), where we have estimated the expectation with the single sample actually seen by the agent in the last line.

Q-learning is guaranteed to converge to the optimal policy in the limit of infinitesimal learning rates and infinite sampling of the state-action space \citep{watkins1992q,sutton2018reinforcement}.
However, following the greedy policy $a^*(s) = \text{argmax}_{a} Q(s, a)$ before convergence of the Q-values can lead to undersampling of the state-action space and poor performance.
It is therefore common to either use an `$\epsilon$-greedy' policy, $\pi(a|s) = \epsilon / |\mathcal{A}| + (1-\epsilon) I_{a = a^*(s)}$, or a softmax-policy, $\pi(a|s) \propto \text{exp}(\beta Q(s, a))$, to collect the experience used to update the Q-values (\Cref{fig:Q}B).
Such exploration strategies and their biological correlates are discussed in more detail in \Cref{sec:scalar_rew}.

These approaches make Q-learning an `off-policy' algorithm, since the policy used in the learning update (the greedy policy) is different from the policy used for action selection (the stochastic policy).
An on-policy alternative known as `SARSA' (state-action-reward-state-action) is also commonly used, where the update rule uses the Q-value corresponding to the action $a_{t+1}$ sampled at the next timestep instead of the greedy action (\Cref{fig:Q}C):
\begin{linenomath*}
\begin{align}
    \Delta  Q(s_t,a_t) \propto - Q(s_t,a_t) + r_t + \gamma Q(s_{t+1}, a_{t+1}).
\end{align}
\end{linenomath*}
This will converge to the true Q-values for a given policy $\pi$, similar to how the TD learning rule in \Cref{eq:TD-learning} converges to the true value function for a given policy, again under assumptions of infinite sampling of the space.

When animals have to choose between actions with different values, studies have found evidence for midbrain dopamine neurons encoding the prediction error used for either Q-learning \citep{roesch2007dopamine,niv2009reinforcement} or SARSA \citep{morris2006midbrain,niv2009reinforcement}.
They may therefore not just be abstract learning algorithms, but instead have plausible implementations in biological neural circuits.
However, the methods considered so far also have notable shortcomings.
For example, the amount of data needed to learn state(-action) values and the assumption of a stationary environment can be prohibitive for animals needing to act in a rapidly changing world, where bad decisions have fatal consequences.

\section{Model-free and model-based reinforcement learning}
\label{sec:model_free_based}

We have so far considered what is known as `model-free' reinforcement learning algorithms.
These involve learning a stimulus-response function that says `when in state $s$, take action $a$'.
Such algorithms do not require much computation at decision time, where they rely on cached state or action values.
However, it can require a lot of experience with the environment to learn these model-free policies, and they can be inflexible in changing environments.
This is incompatible with many aspect of animal behaviour, which we know is adaptive and can benefit from `latent learning' in an environment before a reward-driven task is ever encountered \citep{blodgett1929effect,tolman1948cognitive}.

On the other hand, `model-based' reinforcement learning uses a model of the world to simulate the consequences of different actions at decision time.
This can be much more data efficient, since learning a world model is often easier than learning a full policy (\Cref{fig:MB}A).
In machine learning settings, model-based RL has exhibited impressive performance across a range of tasks with large state spaces, including Atari, chess, shogi, and Go \citep{silver2018general, schrittwieser2020mastering, deisenroth2011pilco}.
In a biological context, the idea of first learning a model of the environment, and then using it to guide reward-driven behaviour also provides one plausible explanation for latent learning and other types of rapid adaptation.
However, model-based decision making can be computationally intensive at decision-time, which is a challenge for animals that rely on rapid decision making for survival (\Cref{fig:MB}B).

In model-based RL, an approximate transition-and-reward function $\tilde{p}(s', r | s, a)$ is learned from past experience.
Once this model has been learned, it can be used for planning at decision time.
This can be done for example by expanding the Q-value relation from \Cref{eq:Q-optimal}:
{\small
\begin{linenomath*}
\begin{align}
    \label{eq:Q-search}
    Q(s_t,a_t) &\approx  r(s_t, a_t) + \gamma \mathbb{E}_{\tilde{p}(s_{t+1} | s_t, a_t)} \left [ \text{argmax}_{a_{t+1}} Q(s_{t+1}, a_{t+1}) \right ]\\
    &\approx r(s_t, a_t) + \gamma \mathbb{E}_{\tilde{p}(s_{t+1} | s_t, a_t)} \left [ \text{argmax}_{a_{t+1}}
    \left [ r(s_{t+1}, a_{t+1}) + \gamma \mathbb{E}_{\tilde{p}(s_{t+2} | s_{t+1}, a_{t+1})} \left [ \text{argmax}_{a_{t+2}} Q(s_{t+2}, a_{t+2}) \right ] \right ] \right ] \\
    &= \ldots
\end{align}
\end{linenomath*}
}

\vspace{-2 em}

If the environment is determinstic, $p(s' | s, a)$ is a delta function, and otherwise the next-state expectations may need to be approximated with multiple samples.
Unfortunately, optimizing over all possible action sequences in \Cref{eq:Q-search} is in general an exponentially large search problem in the planning depth, which makes it infeasible for any reasonably sized problem.
It is therefore common to either use `depth-first search' with limited breadth, or `breadth-first search' with limited depth.
In breadth-first search, we consider all possible actions at each level of the search tree but terminate the search at a finite depth, instead using cached `model-free' state-values to estimate the reward-to-go from the terminal states.
Such `plan-until-habit' has also been proposed as a model of human behaviour \citep{keramati2016adaptive}.
In depth-first search, we instead sample a series of paths from $s_t$ to termination (or some upper bound), using a heuristic to prioritize actions expected to be good, and then pick an action with high expected reward \citep{huys2012bonsai}.

For both of these strategies, it is necessary to trade off the temporal opportunity cost of planning with the increase in expected reward \citep{botvinick2014computational,agrawal2022temporal}.
This has been a popular research area in cognitive science, where a wealth of literature on `resource-rational' decision making has emerged in recent years \citep{griffiths2019doing,callaway2022rational}.
However, this literature has often focused on the behaviour of optimal agents, with less focus on the learning process and neural mechanisms that might implement the necessary computations.
Bridging this gap, recent work has suggested that frontal cortex and striatum might initially store a `model-free' policy in its network state, which is gradually updated with model-based information from the hippocampal formation until the policy improvement is outweighed by the temporal opportunity cost of planning \citep{jensen2023recurrent}.

While several model-based and model-free reinforcement learning methods have thus been developed and used to model animal learning and behaviour, it remains an open question when and whether these different strategies drive animal behaviour.
A different line of research has therefore explicitly investigated the balance between model-based and model-free RL in biological agents \citep{daw2005uncertainty, geerts2020general, lengyel2007hippocampal}, where the choice between the two approaches is thought to be guided by some notion of optimality on the basis of available resources and uncertainty about the environment.
A popular paradigm for these studies is the so-called `two-step' task developed by Daw and colleagues (\citealp{daw2011model,momennejad2017successor,wang2018prefrontal}; although note \citealp{akam2015simple}).

Such work has shown that animals can use both model-free and model-based decision making, with the dorsolateral striatum being particularly important for model-free reinforcement learning \citep{yin2004lesions, yin2005role}, and the dorsomedial striatum, prefrontal cortex, and hippocampal formation being important for model-based decision making \citep{vikbladh2019hippocampal,geerts2020general,miller2017dorsal,niv2009reinforcement,killcross2003coordination}.
This also has interesting parallels to recent work in motor learning, where the basal ganglia were found to be sufficient for `habitual' motor sequences even in the absence of motor cortex, while motor cortex was necessary for more flexible motor behaviours that are likely to require a high-level `schema' of the task structure \citep{mizes2023motor,mizes2023dissociating}.
In \Cref{sec:scalar_rew} we will see how combining these model-based and model-free ideas with deep learning can lead to human-level performance in tasks such as chess and Go that require long-term planning.

\section{The successor representation}
\label{sec:SR}

As we saw in the previous section, an important distinction can be made between model-free reinforcement learning methods, which cache stimulus-response mappings based on prior experience, and model-based reinforcement learning methods, which compute a policy by simulating possible futures using a world model at decision-time.
However, we have also noted how animals both need the flexibility of model-based methods as well as the rapid decision making afforded by model-free methods.
It has therefore been suggested that animals use intermediate methods that combine some model-free and some model-based features.
A particularly prevalent theory has been that of the `successor representation' (SR), which has been proposed to explain both human behaviour \citep{momennejad2017successor} and features of neural activity \citep{stachenfeld2017hippocampus}.
In particular, the SR allows for flexible adaptation to changing reward functions without having to perform expensive simulations at decision time.

\begin{figure*}[!t]
    \centering
    \vspace*{-0.5em}
    \includegraphics[width=0.95\textwidth]{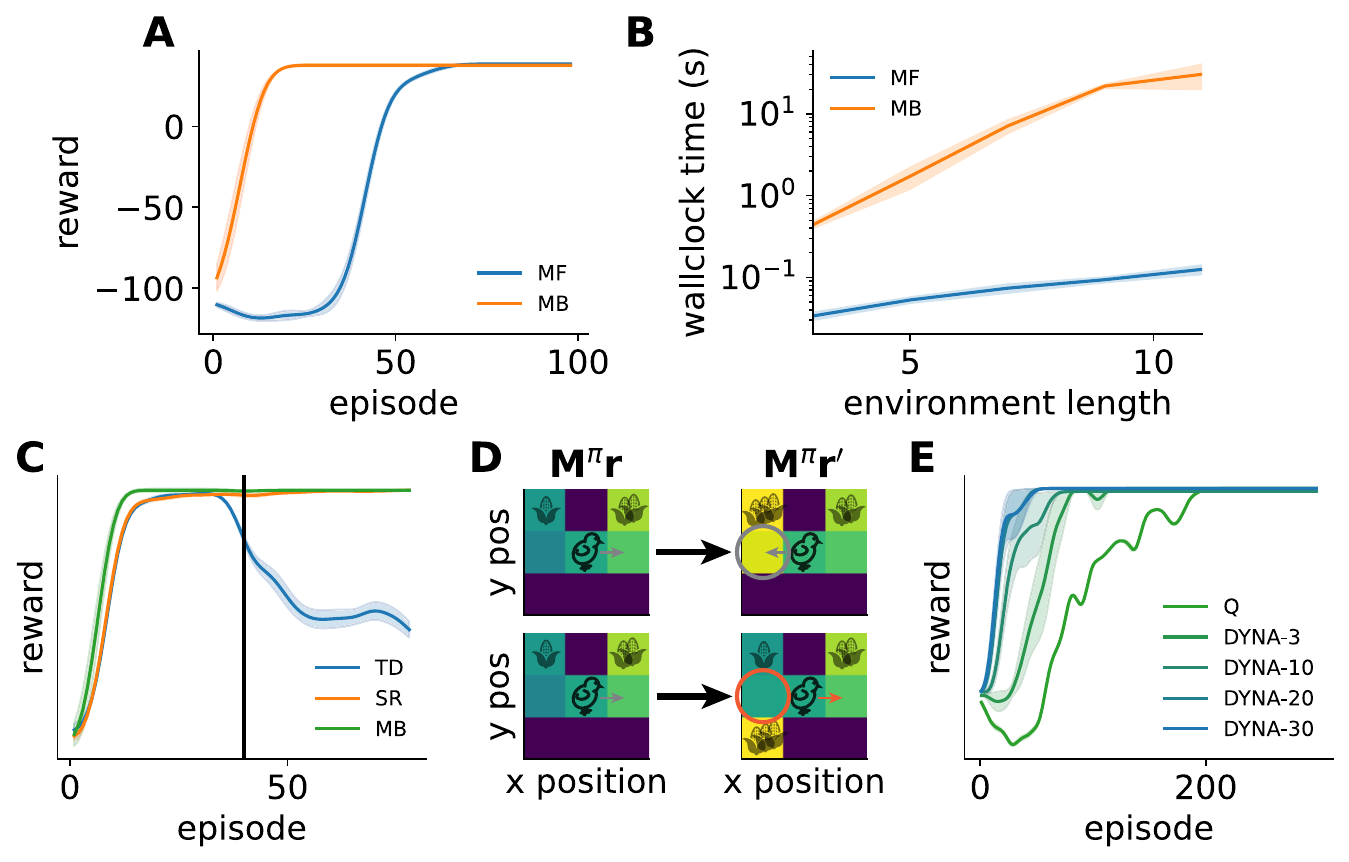}
    \vspace*{-1.0em}
    \caption[MB]{\label{fig:MB} \justifying
        {\bfseries Model-based reinforcement learning.}
        {\bfseries (A)}~Learning curves for model-free (MF) and model-based (MB) RL agents.
        The MB agent used depth-first search to compute an optimal path at each decision point, gradually learning the reward and transition functions while exploring the environment.
        The MF agent was a Q-learning agent with $\epsilon = 0$ and learning rate $\alpha = 1$.
        {\bfseries (B)}~Wallclock time needed to run 100 episodes of cliffworld with either the MB or MF agents from (A), as a function of the length of the environment.
        While the MB agent required less experience to learn a good policy, the wallclock time per episode was much larger than for the MF agent.
        This illustrates an important balance between model-based and model-free reinforcement learning, where MF methods usually require more experience but MB methods require more compute at decision time.
        {\bfseries (C)}~Learning curve for an agent using the successor representation (SR) together with learning curves for the model-based agent in (A) and the greedy TD-agent from \Cref{fig:TD}.
        The goal was moved from location (9, 0) to location (0, 4) at episode 40 (vertical black line), and location (9, 0) was instead given a reward of -5.
        The MB and SR agents had their reward functions updated to reflect this change and rapidly adapted their policies, while the TD agent had no such mechanism for robustness to changing reward functions.
        Reward curves were convolved with a Gaussian kernel ($\sigma = 3$ episodes), which is why performance appears to decrease slightly before episode 40.
        The TD and SR agents were assumed to have access to a 1-step world model at initialization, while the MB agent learned the transition structure from experience.
        {\bfseries (D)}~SR agents cannot always adapt to new reward functions if the newly rewarded states have low probablity under the old policy.
        Left column: Value function for an agent that learned an initial policy in an environment with a small reward in the upper left corner and intermediate reward in the upper right corner.
        The middle top and bottom states are `cliffs'.
        The agent learned to make an initial rightward choice (grey arrows).
        Right column: A large reward was introduced in either the top left corner (top row) or bottom left corner (bottom row) and the value function recomputed (\Cref{eq:SR_recompute}).
        The agent was unable to adapt to a large reward in the bottom left corner, since the old policy had low probability of reaching this state, even after initially going to the left.
        This results in a low expected value for going left from the start state (red circle), and a suboptimal policy that continues to go right (red arrow).
        {\bfseries (E)}~Learning curve for a standard Q-learning agent (blue) or Dyna agents that perform different numbers of Q-value updates after each physical action (legend).
        These Dyna updates used cached experience rather than data from a learned world model.
        Dyna agents make better use of limited experience at the cost of increased compute (proportional to the number of updates).
        }
    \vspace*{-0.5em}
\end{figure*}

The SR \citep{dayan1993improving} rewrites the expected reward starting from state $s$ as:
\begin{linenomath*}
\begin{align}
    V^\pi(s) &= \mathbb{E}_\pi \left [ \sum_{t = 0} \gamma^t r_t | s_0 = s\right ] \\
    &= \sum_{t = 0} \gamma^t \sum_{s'} p_\pi(s_t = s' | s_0 = s) r(s')\\
    &= \sum_{s'} r(s') \sum_t \gamma^t p_\pi(s_t = s' | s_0 = s)\\
    &= \b{r}^T \b{m}^\pi_s.
\end{align}
\end{linenomath*}
Here, $\b{r}$ is a vector of the average reward associated with each state, and $\b{m}^\pi_s$ is a vector of the expected discounted future occupancy of state $s'$ if the agent starts in state $s$ and follows the policy $\pi$:
\begin{linenomath*}
\begin{equation}
    M^\pi_{s s'} = \sum_{t = 0} \gamma^t p_\pi(s_t = s' | s_0 = s).
\end{equation}
\end{linenomath*}
The full matrix $\b{M}^\pi$, constructed from stacking the $\b{m}^\pi_s$ corresponding to all states $s$, is denoted the `successor matrix', and it allows us to write down a vector of expected rewards from all states as
\begin{linenomath*}
\begin{equation}
    \b{v}^\pi = \b{M}^\pi \b{r}.
\end{equation}
\end{linenomath*}
Here, we have retained the superscript $\pi$ to indicate that the successor matrix and value function depend on the policy of the agent, which affects the expected occupancy of different states.
Having computed the value of each state, we can perform action selection using \Cref{eq:value_action_selection}.

The flexibility of the successor representation arises when the reward structure of the environment changes, $\b{r} \rightarrow \b{r}'$.
We can now compute the expected reward associated with each state under the new reward function and old policy,
\begin{linenomath*}
\begin{equation}
    \label{eq:SR_recompute}
    {\b{v}'}^\pi = \b{M}^\pi \b{r}'.
\end{equation}
\end{linenomath*}
This provides a better starting point than the old policy \emph{and} reward function (\Cref{fig:MB}C), but the SR does not generalize perfectly since the new value function can lead to policy changes and therefore having to update $\b{M}$ (\Cref{fig:MB}D).
The successor matrix can be learned by temporal-difference learning when transitioning from $s_t$, analogous to \Cref{eq:TD-learning}:
\begin{linenomath*}
\begin{equation}
    \Delta M^\pi_{s_t s'} \propto -M^\pi_{s_t s'} + I_{s_t = s'} + \gamma M^\pi_{s_{t+1} s'}.
\end{equation}
\end{linenomath*}
${s'}$ is any state, and $s_{t+1}$ is the next state actually observed.
Intuitively, transitioning from $s_t$ to $s_{t+1}$ means that (i) we have just been in state $s_t$, and (ii) we should increase the expected occupancy of all states commonly reached from $s_{t+1}$ (including $s_{t+1}$ itself).
Alternatively, if the policy-dependent transition matrix $\b{T}^\pi$ is known, where $T^\pi_{s s'} = p_{\pi}(s_{t+1} = s' | s_t = s)$, the successor matrix can be computed as the geometric series $\b{M}^\pi = \b{I} + \gamma \b{T}^\pi + \gamma^2 (\b{T}^\pi)^2 + \ldots = (\b{I} - \gamma \b{T}^\pi)^{-1}$.

The SR has been proposed as a model of how humans and other animals learn and generalize \citep{momennejad2017successor, stachenfeld2017hippocampus, geerts2020general,gershman2018successor}.
For example, humans adapt more readily to changes in reward functions than changes in transition functions in a simple sequential binary decision-making task \citep{momennejad2017successor}, consistent with the SR facilitating rapid adaptation to changes in $\b{r}$ but only slow learning of $\b{M}$.
Additionally, hippocampal `place cells' have been proposed to encode a predictive map, with each cell corresponding to a column of $\b{M}$ \citep{stachenfeld2017hippocampus}.
In this model, the firing of a place cell in a given location $s$ reflects the expected future occupancy of its `preferred' location $s'$ conditioned on currently being at $s$.
\citet{stachenfeld2017hippocampus} showed that the SR model explains a range of findings in the hippocampal literature.
For example, this model explains the asymmetry of directional place fields on a one-dimensional track \citep{mehta2000experience}, where $s'$ is more likely to be reached from other states $s$ that \emph{precede} $s'$ than equidistant states that \emph{follow} $s'$.
The SR also explains why place fields change near a newly inserted barrier, since two states on either side of the barrier can no longer be visited in quick succession \citep{alvernhe2011local}.

While the SR is perhaps the most prominent model in systems neuroscience that combines features of model-free and model-based RL, it is not the only one.
Another interesting algorithm is the `Dyna' architecture of \citet{sutton1991dyna}.
In this framework, a model of the world is learned from experience and used to train a model-free policy offline by bootstrapping imagined experience sampled from the model.
This allows for more data-efficient learning of model-free policies at the cost of additional compute during `rest', but without needing more compute at decision time (\Cref{fig:MB}E).
The model used to simulate data for offline training can either be an explicit learned world model, or it can simply be a memory buffer of past experiences in the form of $(s_t, a_t, r_t, s_{t+1}, a_{t+1})$ tuples.
Such experience replay has also proven crucial to the success of modern deep reinforcement learning agents by allowing for higher data efficiency and reducing the instability arising from online experience being autocorrelated \citep{mnih2013playing,schaul2015prioritized}.
A prominent theory in neuroscience posits that hippocampal replays could be implementing such a Dyna-like algorithm by generating imagined experience that is used to train the model-free RL systems of the brain \citep{mattar2018prioritized}.
This theory is supported by the finding that patterns of rodent replay in multiple navigation tasks are consistent with the optimal replays of a Q-learning agent with Dyna, and it has recently been extended to explain not just the content of replays but also their timing \citep{agrawal2022temporal}.

\section{Deep reinforcement learning}
\label{sec:deep_RL}

We have so far considered small state and action spaces, where tabular policies are tractable.
Unfortunately, most ethologically relevant state spaces are large enough that we cannot enumerate all possible states and actions.
However, novel situations often resemble previously encountered states, allowing agents to generalize shared structure to these new but related settings \citep{botvinick2020deep}.
In these cases, we can use \emph{function approximation} \citep{sutton2018reinforcement} instead of tabular policies.
This involves an assumption that similar states will have similar state-action values and should therefore have similar policies.
By making this assumption, we can generalize to unseen states based on previous experience.
The use of deep or recurrent neural networks as powerful function approximators for reinforcement learning has driven impressive progress in this setting -- the domain of `deep reinforcement learning' (deep RL).
Deep RL has seen increasing interest not just in machine learning, but also as a model of neural dynamics and behaviour in humans and other animals \citep{wang2018prefrontal, jensen2023recurrent, makino2023arithmetic, merel2019deep, banino2018vector,aldarondo2024virtual,botvinick2020deep}.
Popular approaches in (model-free) deep RL can largely be divided into two categories: value-based methods, which compute state-action values that can be used for action selection; and policy gradient methods, which train a neural network to output a policy directly.

\subsubsection*{Value-based methods}
\vspace*{-0.0em}

The deep RL approaches most similar to the tabular methods considered in \Cref{sec:temporal_difference} and \Cref{sec:q_learning} use neural networks to compute state-action values, which can be used for action selection as we saw in \Cref{eq:value_action_selection}.
However, by using function approximation instead of the tabular values considered previously, these networks can generalize to unseen states in large state spaces.
This gives rise to the family of `deep Q-learning' methods, which closely mirror the tabular Q-learning considered previously, but now with function approximation.

The simplest approach involves defining a state-action value function $Q_\theta(s, a)$, where the parameters $\theta$ of the neural network defining our agent are learned as follows:
\vspace*{-1.9em}
\begin{itemize}
    \item Collect experience $(s_t, a_t, r_t, s_{t+1})$.
    \item Define a loss $\mathcal{L} = 0.5 [ Q_\theta(s_t, a_t) - (r_t + \gamma \text{max}_a Q_\theta(s_{t+1}, a)) ]^2 $.
    \item Update parameters $\Delta \theta \propto - \frac{\partial \mathcal{L}}{\partial \theta}$ \hspace{0.5em} [often treating the `target value' $y_t = (r_t + \gamma \text{max}_a Q_\theta(s_{t+1}, a))$ as constant w.r.t. $\theta$].
\end{itemize}
\vspace*{-1.3em}
When acting according to our policy, we simply pick the action predicted to have the highest value, usually using some variant of $\epsilon$-greedy or softmax to increase exploration.

On the surface, this looks like a straightforward generalization of tabular Q-learning, and it may seem surprising that deep Q-learning did not see significant use or success until the foundational work of \citet{mnih2013playing}.
However, a major difficulty arises from the autocorrelation of the states observed by the agent, which destabilizes training.
This can be mitigated by the use of `experience replay', where the experience generated by the agent is added to a global replay buffer $\mathcal{B}$.
One or more experiences are then sampled randomly from the buffer at each iteration and used to update the network parameters -- reminiscent of the `Dyna' architecture described previously.
Additionally, the target value $y_t = (r_t + \gamma \text{max}_a Q_\theta(s_{t+1}, a))$ itself depends on $\theta$ and therefore changes when any Q-value is updated (in contrast to tabular Q-learning, where there is no parameter sharing).
It is therefore common to use a `target network' $Q_{\theta'}$ that remains fixed for multiple rounds of data collection and parameter updates.
This reduces fluctuations in the target values, and the resulting parameter updates are gradients of a well-defined objective function.
Together, these two approaches give rise to the `deep Q network' (DQN) developed by \citet{mnih2013playing}, which is trained as follows:
\vspace*{-2.4em}
\begin{itemize}
    \item Collect experience $(s_t, a_t, r_t, s_{t+1})$ and add to $\mathcal{B}$ \hspace{0.5em} [optionally many iterations and optionally removing stale experiences].
    \item Randomly sample an experience $(s'_t, a'_t, r'_t, s'_{t+1}) \sim \mathcal{B}$ \hspace{0.5em} [optionally a full batch].
    \item Define a loss $\mathcal{L}(\theta) = 0.5 [ Q_\theta(s'_t, a'_t) - (r'_t + \gamma \text{max}_a Q_{\theta'}(s'_{t+1}, a)) ]^2 $ \hspace{0.5em} [optionally averaged over the full batch]. Note the `student network' has parameters $\theta$ and the target network inside the max has parameters $\theta'$.
    \item Update the network parameters $\Delta \theta \propto - \frac{\partial \mathcal{L}(\theta)}{\partial \theta}$.
    \item At regular intervels, set our target network to the student network, $\theta' \leftarrow \theta$.
\end{itemize}
\vspace*{-1.3em}
This algorithm is effectively off-policy, since most of the data in $\mathcal{B}$ is collected by a policy defined by an old set of parameters -- and the data in $\mathcal{B}$ can in fact be generated completely independently of the agent being trained.
Even though the DQN is more stable than naive deep Q-learning, an additional instability arises from the fact that $Q_{\text{max}}(s'_{t+1}) = \text{max}_a Q_{\theta'}(s'_{t+1}, a)$ uses the same Q values both to estimate which action is best and what the value of that action is, which leads to a positively biased estimate.
This can be mitigated by `double Q-learning' \citep{van2016deep}, where the student network selects the best action and the target network evaluates its value, $Q_{\text{max}}(s'_{t+1}) \leftarrow Q_{\theta'}(s'_{t+1}, \text{argmax}_a(Q_{\theta}(s'_{t+1}, a)))$.

While modern deep Q-learning has reached impressive performance across a range of machine learning settings \citep{mnih2013playing, lillicrap2015continuous, schaul2015prioritized, kalashnikov2018qt}, it is unclear whether the various modifications needed to stabilize the algorithm could be implemented in biological circuits.
This is perhaps the reason why neuroscience research using deep Q-learning has been relatively scarce, despite the prevalence of tabular Q-learning in theoretical neuroscience.
An interesting exception is recent work by \citet{makino2023arithmetic}, which shows parallels between the values learned by a DQN and neural representations in mammalian cortex during a compositional behavioural task.
Additionally, the importance of experience replay in DQNs \citep{mnih2013playing, schaul2015prioritized} has close parallels to the proposal that hippocampal replay constitutes a form of experience replay \citep{mattar2018prioritized}.

\subsubsection*{Policy gradient methods}
\vspace{-0em}

A conceptually simpler approach for deep reinforcement learning uses policy gradient methods \citep{sutton2018reinforcement}, where a neural network with parameters $\theta$ takes as input the (observable) state of the environment and directly outputs a policy $\pi_\theta$.
This has also found more support and use in the neuroscience literature, where policy gradient methods have recently been used as models of learning and neural dynamics in the biological brain \citep{wang2018prefrontal, jensen2023recurrent, merel2019deep, song2017reward}.

The objective in a policy gradient network is to find the setting of $\theta$ that maximizes expected reward.
A naive way to achieve this would be to define $R_\tau := \sum_{t=0}^T \gamma^t r_t$ and compute gradients given by
\begin{linenomath*}
\begin{align}
    \nabla_\theta J(\theta) &= \nabla_\theta \mathbb{E}_{\tau \sim p_{\pi_\theta}(\tau)} \left [ R_\tau \right ]\\
    &= \sum_{\tau} R_\tau \nabla_\theta p_{\pi_\theta}(\tau).
    \label{eq:simple_deriv}
\end{align}
\end{linenomath*}
Here, $\tau \sim p_{\pi_\theta}(\tau)$ indicates trajectories sampled from the distribution induced by the policy $\pi_\theta$, and $J(\theta)$ indicates the expectation of $R_\tau$ under $p_{\pi_\theta}(\tau)$ (c.f. \Cref{eq:RL_objective}).
However, evaluating \Cref{eq:simple_deriv} requires us to know how the environment will respond to our actions, which in general may not be the case.
Instead, we use the `log-derivative trick', which takes advantage of the linearity of the expectation and the identity $\nabla_\theta \log f(\theta) = f(\theta)^{-1} \nabla_\theta f(\theta)$ to write
\begin{linenomath*}
\begin{align}
    \label{eq:deriv_J}
    \nabla_\theta J(\theta) & = \sum_\tau R_\tau \nabla_\theta p_{\pi_\theta}(\tau) \\
                            & = \sum_\tau R_\tau p_{\pi_\theta}(\tau) \nabla_\theta \log p_{\pi_\theta}(\tau) \\
                            & = \mathbb{E}_{\tau \sim p_{\pi_\theta}(\tau)} \left [ R_\tau \nabla_\theta \log p_{\pi_\theta}(\tau) \right ],
\end{align}
\end{linenomath*}
Since the environment does not depend on $\theta$, we can simplify the calculation of $\nabla_\theta \log p_{\pi_\theta} (\tau)$:
\begin{linenomath*}
\begin{align}
    \label{eq:deriv_log_ptau}
    \nabla_\theta \log p_{\pi_\theta}(\tau) & = \nabla_\theta \left [ \log p(s_0) + \sum_{t=0}^T \log p(s_{t+1} | s_t, a_t) + \log \pi_\theta (a_t|s_t) \right ] \\
                                      & = \sum_{t=0}^T \nabla_\theta \log \pi_\theta (a_t|s_t).
\end{align}
\end{linenomath*}
Inserting \Cref{eq:deriv_log_ptau} in \Cref{eq:deriv_J}, we arrive at the REINFORCE algorithm \citep{williams1992simple}:
\begin{linenomath*}
\begin{align}
    \nabla_\theta J(\theta) & = \mathbb{E}_{\tau \sim p_{\pi_\theta}(\tau)} \left [ R_\tau \sum_{t=0}^T \nabla_\theta \log \pi_\theta (a_t|s_t) \right ]                                         \\
                            & \approx \frac{1}{N} \sum_{\tau \sim p_{\pi_\theta}(\tau)} \left ( \sum_{t=0}^T \gamma^t r_t \right ) \left ( \sum_{t=0}^T \nabla_\theta \log \pi_\theta (a_t|s_t) \right ),
    \label{eq:orig_reinforce}
\end{align}
\end{linenomath*}
where the second line approximates the expectation with $N$ empirical rollouts of the policy in the environment.
Intuitively, \Cref{eq:orig_reinforce} says that we should preferentially upregulate the probability of trajectories with high reward.
Importantly, it no longer differentiates through the environment -- only the policy.

While the REINFORCE algorithm is unbiased, it also has high variance, which can make learning slow and unstable.
It is therefore common to introduce modifications that reduce the variance.
The first of these comes from noting that an action taken at time $t$ cannot affect the reward at times $t'<t$.
We therefore define $R_t := \sum_{t'=t}^T \gamma^{t'-t} r_{t'}$ and write a new update rule as
\begin{linenomath*}
\begin{equation}
    \label{eq:reinforce}
    \hat{\nabla}_\theta J(\theta) \approx \frac{1}{N} \sum_{\tau \sim p_{\pi_\theta}(\tau)}  \sum_{t=0}^T R_t \nabla_\theta \log \pi_\theta (a_t|s_t).
\end{equation}
\end{linenomath*}
This is the formulation most commonly used in the literature, but it is not actually the same as \Cref{eq:orig_reinforce}, which would use $R_t = \sum_{t'=t}^T \gamma^{t'-0} r_{t'}$.
As briefly discussed in \Cref{sec:problem_setting}, this is because the discount factor $\gamma$ is generally used as a variance reduction method rather than because we intrinsically care less about rewards later in the task.
In fact, \Cref{eq:reinforce} is not strictly speaking a gradient \citep{nota2019policy}, which is why we denote it $\hat{\nabla}$.

It can also be shown that subtracting an action-independent baseline from $R_t$ does not change the expectation in \Cref{eq:reinforce}, while potentially reducing its variance.
A common choice is the expected future reward $V(s_t)$:
\begin{linenomath*}
\begin{equation}
    \label{eq:AC}
    \hat{\nabla}_\theta J(\theta) \approx \frac{1}{N} \sum_{\tau \sim p_{\pi_\theta}(\tau)}  \sum_{t=0}^T (R_t - V(s_t)) \nabla_\theta \log \pi_\theta (a_t|s_t).
\end{equation}
\end{linenomath*}
Intuitively, \Cref{eq:AC} upregulates the probability of actions that lead to higher-than-expected reward and downregulates the probability of actions that lead to lower-than-expected reward.

Finally, it is common to reduce the variance of the gradient estimate further through an approach known as `bootstrapping', which approximates $R_t \approx r_t + \gamma V(s_{t+1})$.
This is useful because $r_t + \gamma V(s_{t+1})$ has lower variance than $\sum_{t'=t}^T \gamma^{t'-t} r_{t'}$.
We therefore replace $(R_t - V(s_t))$ in \Cref{eq:AC} with the `advantage function' $A(s_t, a_t) = Q(s_t, a_t) - V(s_t) \approx r_t + \gamma V(s_{t+1}) - V(s_t)$.
In between these two extreme cases of a full Monte Carlo estimate of $R_t$ and a `one-step' bootstrap, the sum in $R_t$ can be truncated to any order, with $R_{t'}$ replaced by $V(s_{t'})$ \citep{sutton2018reinforcement}.
In theory, this gradient estimate remains unbiased if the value function is correct.
In practice, the learned estimate of $V(s_{t'})$ will be inexact, which biases the parameter updates.
Bootstrapping therefore leads to a tradeoff between the bias and variance of parameter updates.

These variance reduction approaches give rise to the so-called `actor-critic' algorithm, where an agent both computes a policy $\pi$ (the actor) and an `evaluation' of the policy in the form of state(-action) values (the critic).
A rich neuroscience literature suggests that the basal ganglia of biological agents implement an actor-critic-like algorithm.
Here, dorsal striatum is proposed to implement the `actor' and ventral striatum the `critic' \citep{takahashi2008silencing,sutton2018reinforcement,o2004dissociable}.

For these actor-critic algorithms, it is common to parameterize both the policy $\pi_\theta(a|s)$ and value function $V_\theta(s)$ with neural networks.
To optimize these parameters using out-of-the-box automatic differentiation, we need to write down an `objective function' with the correct gradients -- but we saw in \Cref{eq:simple_deriv} that this cannot simply be the expected reward.
Instead, we define an auxiliary utility (i.e. negative loss)
\begin{linenomath*}
\begin{equation}
    \label{eq:Jtilde}
    \tilde{J}(\theta) = \frac{1}{N} \sum_{\tau \sim p_{\pi_\theta}(\tau)}  \sum_{t=0}^T (R_t - V(s_t)) \log \pi_\theta (a_t|s_t),
\end{equation}
\end{linenomath*}
where $R_t$ can optionally be approximated by $r_t + \gamma V(s_{t+1})$.
While $\tilde{J}(\theta)$ has no intrinsic interpretation, it is chosen such that $\nabla_\theta \tilde{J}(\theta) = \hat{\nabla}_\theta J(\theta)$ when treating $\delta_t := (R_t - V(s_t))$ as constant w.r.t $\theta$, and the gradients can be computed using standard automatic differentiation.
The gradient of the value function loss is then given by $\nabla_\theta \sum_t \frac12  (R_t - V_\theta(s_t))^2 = \sum_t \left [ - \delta_t \nabla_\theta V_\theta(s_t) \right ]$.

While these policy gradient methods may seem far removed from neuroscience, it has been found that neural networks trained with policy gradients often learn representations and behaviours reminiscent of biological organisms \citep{wang2018prefrontal, jensen2023recurrent, merel2019deep,li2022integrating,song2017reward}.

\subsubsection*{Meta-reinforcement learning}

\begin{figure*}[!t]
    \centering
    \vspace*{-0.5em}
    \includegraphics[width=0.95\textwidth]{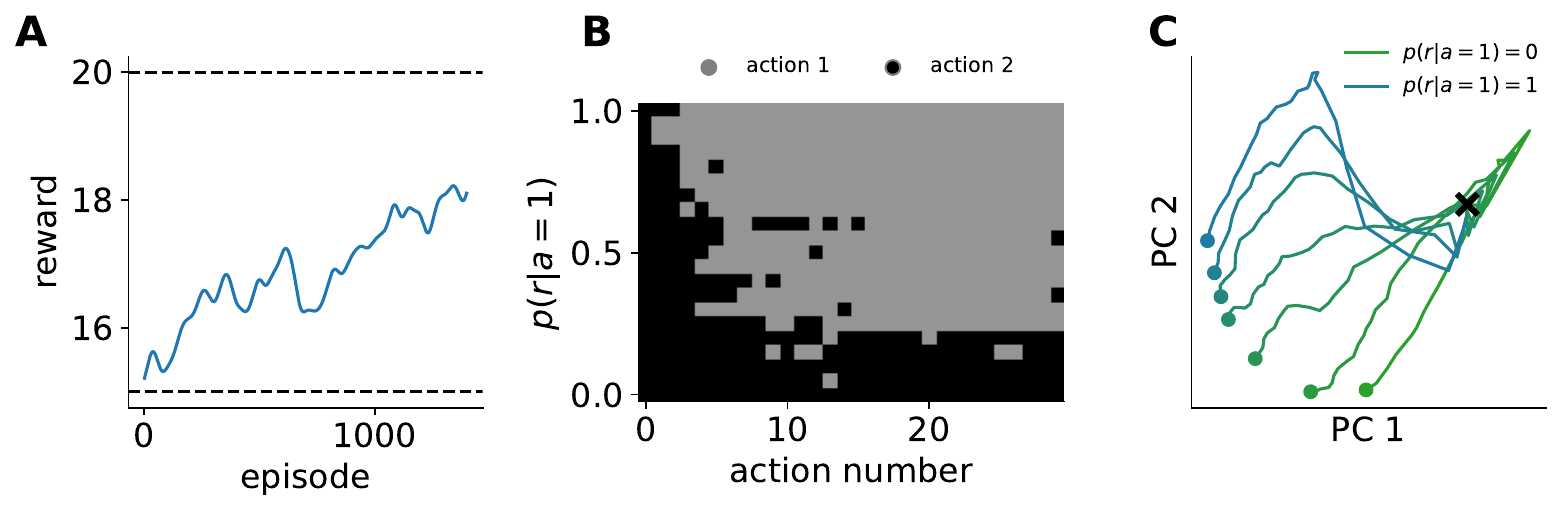}
    \vspace*{-1.0em}
    \caption[metaRL]{\label{fig:meta} \justifying
        {\bfseries Meta-reinforcement learning.}
        The results in this figure reproduce some of the analyses in Figure 1 of \citet{wang2018prefrontal}.
        {\bfseries (A)}~We trained a recurrent meta-reinforcement learning agent in a two-armed bandit task, where the reward probabilities of each arm were sampled independently from $\mathcal{U}(0, 1)$ at the beginning of each episode and remained fixed throughout the episode.
        A recurrent neural network was trained across many episodes with \emph{different reward probabilities} using an actor-critic algorithm.
        The input to the agent consisted of the previous action, the previous reward, and the time-within-trial.
        The average reward per episode is plotted against the episode number, showing that the agent gradually learns to adapt within each episode to the particular instantiation of the bandit task.
        Importantly, the parameters of the network are fixed within an episode, meaning that this adaptation occurs through the recurrent dynamics.
        Dashed horizontal lines indicate the reward of an agent selecting random actions and an `oracle' agent that always chooses the best arm.
        {\bfseries (B)}~Heatmap showing example behaviour of the agent in episodes with different reward probabilities for the first arm, $p(r | a = 1)$.
        For the analysis here and in (C), we set $p(r | a = 2) = 1 - p(r | a = 1)$.
        Across episodes, the agent experiments with different actions and eventually converges on the optimal action.
        For episodes with more similar reward probabilities (near the middle), it takes longer to identify the optimal action.
        This balance between exploration and exploitation is mediated by the recurrent network dynamics, which are learned over many episodes using deep reinforcement learning.
        {\bfseries (C)}~We averaged the hidden state of the RNN over 100 episodes for each of several different reward probabilities, ranging from low (green) to high (blue) $p(r | a = 1)$.
        We then performed PCA on the resulting matrix of average hidden states to compute a low-dimensional trajectory over the course of an episode for each reward probability.
        This two-dimensional embedding of neural activity converges to different regions of state space during the episode for different reward probabilities.
        Black cross indicates the hidden state at the beginning of an episode, and coloured points indicate the final hidden state in an episode for the different reward probabilities.
        }
    \vspace*{-0.5em}
\end{figure*}

A prominent example of deep reinforcement learning providing insights into biological circuits is the `recurrent meta-reinforcement learning' model developed by \citet{wang2018prefrontal}.
The authors trained a \emph{recurrent} deep RL agent using policy gradients, where the RNN parameters are configured by learning from rewards over long periods of time from many tasks that have a shared underlying structure.
Importantly, this `slow' model-free learning process gives rise to an agent that can rapidly learn from experience with \emph{fixed parameters} when exposed to a new task from the same task distribution.
This is achieved by the agent learning to effectively implement a fast RL-like algorithm in the \emph{dynamics} of the network (\Cref{fig:meta}).
This process, whereby an agent trained slowly on a large distribution of tasks can rapidly adapt to a new task, is known as `meta-reinforcement learning' \citep{finn2017model, ritter2018been, duan2016rl, wang2016learning}.
\citet{wang2018prefrontal} suggested that prefrontal cortex resembles such a recurrent meta-RL system, and their model explained a range of neuroscientific findings.
This included
\vspace*{-1.9em}
\begin{itemize}
    \item Dynamic adaptation of the effective learning rate of an agent to the volatility of the environment \citep{behrens2007learning}.
    \item The emergence of `model-based' behaviour in the `two-step' task commonly used to distinguish between model-free and model-based RL \citep{miller2017dorsal,daw2011model}.
    \item The ability of animals to get progressively faster at learning when exposed to multiple tasks with a consistent abstract task structure \citep{harlow1949formation}.
\end{itemize}
\vspace*{-1.1em}
Further experimental evidence for this meta-RL framework comes from \citet{hattori2023meta}, who showed that across-session learning in a reversal learning task relied on synaptic plasticity in orbitofrontal cortex (OFC; a subregion of PFC), while within-session learning relied on recurrent dynamics in OFC.
Recently, \citet{jensen2023recurrent} also extended the work of \citet{wang2018prefrontal} to allow the meta-RL network dynamics to update the policy from \emph{imagined} experience using a learned model of the environment -- reminiscent of Dyna, but now implemented in RNN dynamics instead of parameter updates.

\section{Distributional reinforcement learning}
\label{sec:distributional}

\begin{figure*}[!t]
    \centering
    \vspace*{-0.5em}
    \includegraphics[width=1.0\textwidth]{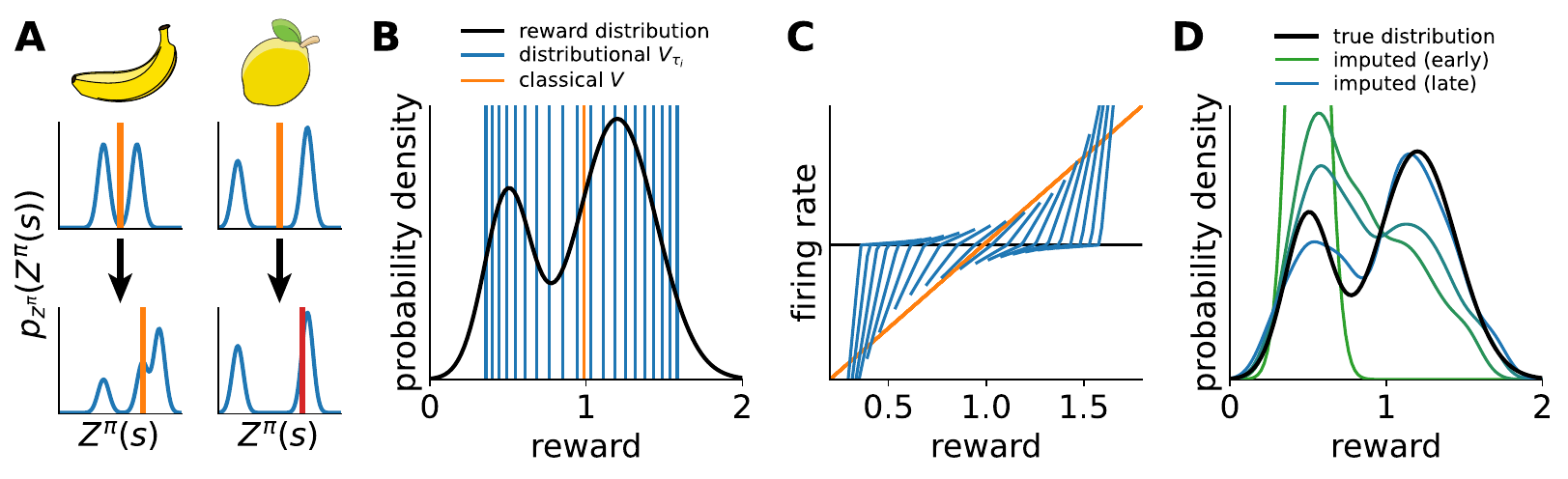}
    \vspace*{-2.5em}
    \caption[Distributional]{\label{fig:dist} \justifying
        {\bfseries Distributional reinforcement learning.}
        {\bfseries (A)}~Example of distributional RL improving representation learning.
        Two states (e.g. `banana' and `lemon') may have the same expected future reward but different reward \emph{distributions} (top row).
        A standard RL agent only has to predict the mean (orange vertical lines) and may learn a simple predictive feature like `yellow' that combines both states.
        If the expected future reward now changes for one state (bottom row), e.g. because the agent learned to make a banana smoothie, it may erroneously generalize to \emph{all} yellow fruits (bottom right; red line).
        However, a distributional agent is forced to learn an initial representation that distinguishes the states, which can improve downstream learning and prevent overgeneralization (blue curves).
        {\bfseries (B)}~Distributional RL simulations on a value estimation task with a single state and no actions, reproducing key ideas from \citet{dabney2020distributional}.
        Distribution of rewards (black), plotted together with the learned values $V_{\tau_i}$ across 20 units learning through standard TD learning (orange; all $\tau_i = 0.5$), or through distributional RL with different $\tau_i = \frac{\alpha_i^+}{\alpha_i^+ + \alpha_i^-}$ (blue).
        The TD units all converge to the mean reward, while the expectile units end up tiling the distribution.
        {\bfseries (C)}~For both the TD units and distributional units from (B), we plot the temporal difference updates performed in response to different rewards from the reward distribution.
        These updates have been proposed to be represented in the firing rates of dopaminergic VTA neurons relative to baseline \citep{schultz1997neural,dabney2020distributional}.
        The TD units show a constant linear scaling across positive and negative rewards, while the distributional units show an asymmetric scaling of firing rate with reward magnitude above and below their reversal point (black horizontal line).
        The ratio of slopes above and below the reversal point scales positively with the value of the reversal point.
        These features of dopaminergic VTA neurons were used by \citet{dabney2020distributional} to argue that the brain implements a form of distributional RL.
        {\bfseries (D)}~True reward distribution (black) reproduced from (B), now plotted together with the reward distribution $p_{\{ V_{\tau_i} \} }(\tilde{Z}^\pi(s))$ implied by the distributional units from (B) and (C) at different stages of learning (green to blue).
        These imputed distributions were computed by \emph{assuming} that the value $V_{\tau_i}$ learned by unit $i$ corresponds to expectile $\tau_i = \frac{\alpha_i^+}{\alpha_i^+ + \alpha_i^-}$ of the reward distribution.
        We then infer the distribution implied by these expectiles under the assumption that it consists of a set of 20 delta functions \citep{rowland2019statistics}.
        Finally, we convolved the resulting delta functions with a Gaussian kernel ($\sigma = 0.1$) for visualization.
        This process was repeated using $\{V_{\tau_i}\}$ at different stages of learning.
        The units were all initialized at $V_{\tau_i} = 0.5$, so the initial distribution is a delta function at $r = 0.5$ (green).
        At the end of learning, the population faithfully represents the true reward distribution, capturing key features including bimodality and the relative magnitude of the two modes (blue).
        \citet{dabney2020distributional} used a similar approach to infer the distribution encoded by dopaminergic VTA neurons at the end of animal training and found a close match to the true reward distribution.
        }
    \vspace*{-0.5em}
\end{figure*}

In \Cref{eq:V-values} and \Cref{eq:Q-values}, we defined the expected future reward for a given state or state-action pair.
The methods considered so far have only used such expectations as a learning signal.
However, recent research suggests that animals may in fact estimate entire future reward distributions \citep{dabney2020distributional, sousa2023dopamine}.
These studies were inspired by findings that such \emph{distributional RL} can improve the performance of artificial agents \citep{bellemare2017distributional, bellemare2023distributional,dabney2018distributional}.
To formalize this, we use $Z^\pi(s, a)$ to denote a single sample from the distribution over possible cumulative discounted future rewards resulting from following policy $\pi$ after taking action $a$ in state $s$:
\begin{linenomath*}
\begin{equation}
    Z^\pi(s, a) \sim p_{z^\pi} \left ( \sum_{t' >= t} \gamma^{t' - t} r_{t'} | s_t = s, a_t = a \right )
\end{equation}
\end{linenomath*}
The stochasticity of $Z^\pi$ can both be due to stochasticity in environment dynamics and reward, and it can be due to stochasticity in the policy of the agent itself.
Clearly, the expectation of $Z^\pi(s, a)$ equals the corresponding Q value:
\begin{linenomath*}
\begin{equation}
    \mathbb{E}_{p_{z^\pi}} \left [ Z^\pi(s, a) \right ] = Q^\pi(s, a).
\end{equation}
\end{linenomath*}
Instead of only estimating this expectation, we now want to learn the full distribution of returns, $p_{z^\pi}(Z^\pi(s, a))$.
One normative reason to learn this distribution is to develop methods that are risk averse \citep{morimura2010nonparametric,morimura2012parametric} or explicitly take into account uncertainty \citep{dearden1998bayesian}.
However, recent work has suggested that such a distributional approach can also increase expected reward by improving representation learning in the deep RL setting \citep{bellemare2017distributional,dabney2018distributional,rowland2019statistics,bellemare2023distributional}.
This is because traditional deep RL only distinguishes states that have different expected value, while distributional RL learns to distinguish states that have different value distributions (\Cref{fig:dist}A).

To implement distributional RL, we consider the $\tau\textsuperscript{th}$ \emph{expectile} of $p_{z^\pi}(Z^\pi(s, a))$, $\epsilon_\tau$, which is defined for a random variable $Z$ as the solution to the equation
\begin{linenomath*}
\begin{equation}
    \tau \mathbb{E} [\text{max}(0, Z - \epsilon_\tau)] = (1-\tau) \mathbb{E} [\text{max}(0, \epsilon_\tau - Z)].
\end{equation}
\end{linenomath*}
This is a generalization of the mean, which is recovered for $\tau = 0.5$, similar to how the quantile generalizes the notion of a median.
A distribution is uniquely specified by its expectiles, and we can therefore represent $p_{z^\pi}(Z^\pi(s, a))$ in terms of $\{\epsilon_\tau\}$.
Translating this to an RL algorithm involves training a network (or tabular values) to predict a set of expectiles for a given state (and action).
The parameters of the agent are then updated by propagating the distribution implied by the predicted expectiles through the Bellman equation and minimizing the difference between the initial and propagated distributions \citep{bellemare2023distributional}.

In the tabular value learning case (c.f. \Cref{sec:temporal_difference}), this can be achieved using a modified TD-update rule (c.f. \Cref{eq:TD-learning}; \citealp{lowet2020distributional}; \Cref{fig:dist}B).
In particular, we consider a set of units $\{ V_{\tau_i}(s) \}$, each with a target expectile $\tau_i := \frac{\alpha_i^+}{\alpha_i^+ + \alpha_i^-}$ of the return distribution $p_{z^\pi}(Z^\pi(s))$ (where $p_{z^\pi}(Z^\pi(s)) = \sum_a \left [ \pi(a|s) p_{z^\pi}(Z^\pi(s)) \right ] $).
These expectiles can be learned by sampling experience from the environment under the policy and defining a TD error for each unit and state transition as
\begin{linenomath*}
\begin{align}
    \delta_i := r_t + \gamma \tilde{Z}^\pi_j(s_{t+1}) - V_{\tau_i}(s_t).
\end{align}
\end{linenomath*}
Here, $\tilde{Z}^\pi_j(s_{t+1})$ is a \emph{random sample} from the learned approximate distribution of cumulative future returns from state $s_{t+1}$, $p_{\{ V_{\tau_i} \} }(\tilde{Z}^\pi(s_{t+1}))$ \citep{lowet2020distributional,dabney2020distributional}.
We then apply the following update rule to all units:
\begin{linenomath*}
\begin{equation}
    \label{eq:DRL_V_expec}
    \Delta V_{\tau_i}(s_t) = \alpha_i^+ \text{max}(\delta_i, 0 ) + \alpha_i^- \text{min}(\delta_i, 0).
\end{equation}
\end{linenomath*}
In other words, we apply the TD update rule to each unit with learning rate $\alpha_i^+$ for positive TD errors and learning rate $\alpha_i^-$ for negative TD errors.
When running this algorithm to convergence, $V_{\tau_i}(s)$ will approach the $\tau_i\textsuperscript{th}$ expectile ($\epsilon_{\tau_i}$) of $p_{z^\pi}(Z^\pi(s, a))$.
In the deep RL setting, we would compute gradients of the form $\nabla_\theta \mathcal{L} = \sum_i \Delta V_{\tau_i}(s_t) \partial V_{\tau_i}(s_t) / \partial \theta$ to learn a model with parameters $\theta$ that predicts the full set of expectiles.
We refer to \citet{bellemare2017distributional,dabney2018distributional,rowland2019statistics,bellemare2023distributional}; and \citet{dabney2020distributional} for additional mathematical details, alternative parameterizations of the return distribution, and extensions to the control setting.

Intriguingly, recent work in neuroscience suggests that distributional RL could underlie value learning in biological neural circuits \citep{dabney2020distributional,lowet2020distributional,sousa2023dopamine}.
In particular, \citet{dabney2020distributional} recorded the activity of dopaminergic VTA neurons during a task with stochastic rewards and showed that the neurons appeared to represent the full distribution of possible outcomes using an expectile representation.
More concretely, they showed that:
\vspace*{-2.1em}
\begin{itemize}
    \item The VTA neurons exhibited a range of different `reversal points' -- defined as the reward magnitude at which the firing rate of a neuron did not change from its baseline firing rate.
    This is consistent with a distributional RL theory, where the changes in neural firing rates from baseline correspond to the expectile TD updates considered above.
    In this case, the reversal point of a neuron $i$ should be $V_{\tau_i} \approx \epsilon_{\tau_i}$ (\Cref{fig:dist}C).
    \item Neurons had different slopes describing the relationship between expected reward and firing rate in the regimes where expected reward was above and below the reversal point ($V_{\tau_i}$) of each neuron.
    This is consistent with the algorithm described in \Cref{eq:DRL_V_expec} (\Cref{fig:dist}C).
    \item When independently fitting a slope to the data above ($\alpha_i^+$) and below ($\alpha_i^-$) the reversal point of neuron $i$, the reversal point of the neuron was positively correlated with $\tau_i = \frac{\alpha_i^{+}}{\alpha_i^- + \alpha_i^+}$.
    This is consistent with the expectile distributional RL setting, where the reversal point is $V_{\tau_i} \approx \epsilon_{\tau_i}$, which is monotonic in $\tau_i$ (\Cref{fig:dist}C).
    \item When `imputing' the distribution implied by the VTA neurons when interpreted as expectiles (\Cref{fig:dist}D), the resulting fit resembled the true distribution of rewards in the experiment.
\end{itemize}
\vspace*{-1.1em}
These findings generalize the canonical RPE view of \citet{schultz1997neural}, which can be seen as the averaged version of the theory by \citet{dabney2020distributional}.
The expectile regression algorithm investigated by \citet{dabney2020distributional} relies on non-local TD updates and non-linear `imputation' of the return distribution $p_{\{ V_{\tau_i} \} }(\tilde{Z}^\pi(s))$ induced by the learned expectiles $\{ V_{\tau_i}(s) \}$. 
However, recent work has suggested more biologically plausible distributional RL updates to address these challenges \citep{tano2020local,wenliang2023distributional}.

\section{Learning from scalar rewards?}
\label{sec:scalar_rew}

The methods discussed so far have all relied on some scalar `reward' available from the environment.
Distributional RL predicted the full distribution of possible rewards but still assumed that an external reward was eventually observed.
This is perhaps reasonable in many experimentally controlled settings, where we deliver juice or water to animals as a function of their actions.
It is also often appropriate in machine learning, where we want to optimize some externally imposed objective that can be quantified.
However, in natural environments, there is no such `global' reward -- instead, different actions can be rewarded in different ways that must be converted to an internal learning signal \citep{juechems2019does}.
For example, foraging for food might reduce hunger, while collecting water might reduce thirst.
Starting a family might have an initial energy cost, but it could improve survival later in life and propagate the gene pool.
An RL purist would convert all of these different gains and losses into a single common unit and balance them appropriately to decide what actions to take \citep{silver2021reward}.
However, it has also been suggested that biological agents adaptively change their instantaneous objective via a higher-order controller in e.g. prefrontal cortex, which determines what the current lower-level objective is \citep{juechems2019does,miller2001integrative,botvinick2008hierarchical}.
This resembles ideas in hierarchical reinforcement learning \citep{pateria2021hierarchical,botvinick2008hierarchical}, but here the higher-order policies are not necessarily learned through reinforcement learning within a lifetime, instead emerging during evolution where survival depends on the discovery of `useful' objectives.

A related challenge is the technical difficulty of learning from scalar rewards, which are generally sparse.
While the methods in \Cref{sec:deep_RL} can train a neural network to maximize reward in theory, in practice they are often noisy, unstable, or find local minima.
As a topical example, consider the challenge of training large language models to interact with human users.
Training such a model from scratch using reinforcement learning would be near-impossible, but `pretraining' the model on a large unsupervised dataset followed by `reinforcement learning from human feedback' has proven hugely successful and revolutionized language models for human interactions \citep{team2023gemini}.
This works because the large unlabelled dataset provides a way to learn good \emph{representations} that distinguish different concepts, since this is necessary to solve the base task of predicting the next word.
Once such representations are learned, the subsequent finetuning for human interaction is an easier problem that can be solved with reinforcement learning.

Of more relevance to neuroscience, it is also common in deep RL to introduce additional `auxiliary costs' to the utility function that are jointly optimized together with reward maximization, and which can use a richer data source to drive representation learning in the network \citep{jaderberg2016reinforcement}.
A popular approach is to include losses that require the agent to predict the next observation from the current state and action \citep{jaderberg2016reinforcement, zintgraf2019varibad}.
Such predictive losses have close parallels in neuroscience, where it has been suggested that predictive learning could drive many of the representations observed in biological circuits \citep{rao1999predictive, stachenfeld2017hippocampus, whittington2020tolman, blanco2021dopamine} and serve as a foundation for model-based planning \citep{jensen2023recurrent}.
\citet{fang2023predictive} also recently showed that augmenting an RL agent with an auxiliary predictive objective leads to neural representations that resemble biological circuits more closely.
This suggests a potentially important interaction between self-supervised representation learning and reward-driven reinforcement learning in biological circuits.

If a good predictive model of the world is learned, this model can also be used together with a search algorithm to turn the reinforcement learning problem into a supervised learning problem.
As an example, the MuZero model developed by \citet{schrittwieser2020mastering} to learn Atari, chess, shogi, and Go, was trained to predict future values and actions by unrolling the environment using a learned latent representation.
It then used Monte Carlo tree search (MCTS; a model based-search algorithm c.f. \Cref{sec:model_free_based}) and the predicted value function to improve the provisional actions predicted by the base policy network.
These final actions, which had been optimized using MCTS, were treated as a supervised learning target for the base policy network.
Through many iterations of such predictive learning, MCTS-based policy improvement, and training of the base policy, the network learned both the transition structure of the task and to select good actions.
Such semi-supervised approaches to reinforcement learning are useful in the common setting where learning the transition structure of the world is easier than learning a policy \citep{jensen2023recurrent}.
These methods also have interesting parallels to learning in biological networks, where interactions between model-based and model-free systems are similarly thought to drive action selection based on learned latent representations \citep{botvinick2020deep}.

Finally, algorithms that seek to maximize scalar rewards often run into challenges related to exploration.
In particular, once an above-average policy has been identified, exploration is disfavored because it has lower expected reward, even if there is a finite probability of discovering new and better policies.
This is why imposing stochasticity in the form of e.g. $\epsilon$-greedy policies is common in the tabular RL literature.
In deep RL, it is instead popular to include various forms of `exploration bonuses' in the objective function.
Policy gradient algorithms for example often add an auxiliary entropy loss of the form $\mathcal{L}_E = \sum_{\tau \sim p_{\pi_\theta}(\tau)} \sum_{t=0}^T \sum_a \pi_\theta (a_t|s_t) \log \pi_\theta (a_t|s_t)$, and indeed the bandit example in \Cref{fig:meta} does not work without such an entropy loss.
Other approaches to improve exploration include hierarchical reinforcement learning, which introduces autocorrelations in the policy to improve coverage of the state space.
This is reminiscent of the biological `Lévy flight' hypothesis, which suggests that animals explore an environment using a heavy-tailed distribution of `step sizes' to maximize the probability of finding sparse rewards \citep{viswanathan1999optimizing}.
Biological agents are also thought to engage in periods of `random exploration' that can be interpreted as a biological parallel to $\epsilon$-greedy-like algorithms.
Random exploration appears to be under noradrenergic control \citep{tervo2014behavioral, dubois2021human}, while the zona incerta can drive more directed exploration \citep{ahmadlou2021cell}.
Such neural control of exploration is particularly well-characterized in the zebra finch song circuit \citep{olveczky2005vocal}, and it is consistent with a view of top-down imposition of flexible reward functions -- where one possible objective could be to increase variability or reduce uncertainty -- rather than optimization of a global scalar reward.

\section{Discussion}
\label{sec:discussion}

In this review, we have provided a mathematical overview of some of the many reinforcement learning methods that are commonly used in systems and computational neuroscience.
We have also highlighted a range of explicit parallels between these methods and experimental results in neuroscience and cognitive science to illustrate the utility of reinforcement learning as a framework for understanding biological learning and decision making.
This has ranged from classical work on reward prediction errors \citep{schultz1997neural} to recent findings of distributional reinforcement learning in biological circuits \citep{dabney2020distributional}.

While RL has already had a profound influence on systems neuroscience, several open questions remain.
In particular, much work has focused on simple stimulus-response or binary decision making tasks.
This is a far reach from ethologically relevant problems that involve processing multimodal stimuli, decision making with long-lasting consequences, and high-dimensional motor control.
Some recent work building on deep RL has started to bridge this gap.
For example, \citet{banino2018vector} demonstrated the emergence of grid cells in agents navigating complex environments, \citet{aldarondo2024virtual} showed similarities between an RL agent trained to control a virtual rodent and corresponding biological motor representations, and \citet{jensen2023recurrent} showed parallels between a recurrent meta-RL agent and human behaviour in a navigation task requiring temporally extended thinking.
However, much work remains to extend our neuroscientific understanding to ethologically relevant settings, both experimental and computational.

A related challenge will be to combine different components of existing models to capture the generalist nature of biological circuits.
This is in contrast to past work, which has often focused on a single neural circuit or function, such as motor control or navigation.
Such a generalist approach will involve explicit modeling of the roles of different brain regions, and more importantly it will require us to capture how they interact with one another during learning and decision making.
Clearly, such models will need to be constrained by experimental data, both at the level of behaviour and at the level of neural activity.
This is becoming increasingly feasible with recent advances in recording technologies, both for high-resolution behavioural tracking \citep{mathis2018deeplabcut,dunn2021geometric} and for simultaneous and long-term recording of neural activity \citep{steinmetz2021neuropixels, pachitariu2016suite2p, dhawale2017automated}.

Finally, most work on reinforcement learning in neuroscience has considered short-term decision making tasks, where planning and decision making in primitive state and action spaces are feasible.
This is in stark contrast to most human decision making, which occurs over extended timescales and often involves hierarchies of decisions or priorities that change over time.
For example, we may decide to pursue an undergraduate degree at Cambridge University, which then requires us to (i) write an application, (ii) prepare for an interview, and (iii) arrange our travel.
Each of these processes in turn require us to plan increasingly low-level decisions, such as booking a flight or deciding which bus to take to the airport.
This is the topic of hierarchical reinforcement learning, which has already been highlighted as a potentially useful model of human behaviour \citep{eckstein2020computational,botvinick2008hierarchical,botvinick2009hierarchically} and is becoming an increasingly important area of research in machine learning \citep{pateria2021hierarchical}.
While reinforcement learning will undoubtedly remain important for understanding such flexible animal behaviour, we should also keep in mind the shortcomings and challenges of learning from scalar rewards (\Cref{sec:scalar_rew}).
It will be important to investigate the regimes in which reinforcement learning provides a good model of behaviour and neural dynamics, while also exploring other frameworks with potentially richer learning substrates.
This will also enable the study of interactions between such different learning algorithms, which will likely be necessary to understand biological learning in rich environments.

\section*{Acknowledgements}
Guillaume Hennequin, Will Dorrell, and the NBDT reviewers and editor provided valuable feedback that improved and clarified the manuscript.
Most of the content in this review was originally prepared for the 2023 Janelia Theoretical Neuroscience Workshop.

\bibliography{references}

\newpage

\section{Additional topics of interest}
\label{sec:additional}

While we have tried to provide a fairly comprehensive overview of topics in reinforcement learning of interest to neuroscience, there are naturally many interesting areas that we have had to omit.
Here we provide a brief description of some of these together with pointers to relevant literature for those who are interested in exploring them further.

\subsection{Hierarchical reinforcement learning}
\label{sec:HRL}
So far, we have considered a simple environment consisting of discrete states and actions, and all planning and decision making has taken place in the space of action primitives.
However, when planning over longer horizons, it can be necessary to break down the overall policy into a series of sub-goals, sub-policies, or `skills' \citep{sutton1999between, pateria2021hierarchical}.
This is the topic of hierarhical reinforcement learning (HRL) and `options', where an agent learns a high-level policy over policies that can themselves be specified in terms of primitive actions or even lower-level policies.
Such HRL has been found to explain features of human behaviour \citep{eckstein2020computational,botvinick2008hierarchical,botvinick2009hierarchically} and remains an area of substantial interest in the neuroscience literature.

\subsection{Off-policy \& offline reinforcement learning}
\label{sec:off-policy}
In most of the work considered in this paper, the experience used to train the RL agents has been sampled from the policy of the agent itself.
Indeed this is required for the gradients to be unbiased in the vanilla policy gradient setting.
However, an area of substantial interest is that of offline reinforcement learning, where the agent is trained from scratch on the basis of pre-collected experience \citep{levine2020offline}.
This is particularly important in cases where online data collection is expensive or too risky but large-scale datasets exist, such as in many healthcare settings.
Off-policy reinforcement learning is the related problem of learning from a combination of online data and pre-generated data, possibly from a `stale' version of the current agent.
The off-policy setting is especially relevant to biology, where data collection is expensive and we therefore wish to make maximum use of existing data.
This can e.g. be achieved through experience replay, which can be prioritized (instead of sampled at random) to maximize future reward and minimize temporal opportunity costs \citep{mattar2018prioritized, agrawal2022temporal,schaul2015prioritized}.
A variety of `off-policy' policy gradient methods have also been developed to improve sample efficiency, which de-bias the gradients e.g. through the use of importance sampling \citep{espeholt2018impala,jie2010connection,peshkin2002learning,haarnoja2018soft}.

\subsection{Imitation learning}
\label{sec:imitation}
Related to the problem of offline reinforcement learning is that of \emph{imitation learning}, where we also learn from pre-collected data.
However, in contrast to offline RL where we make no assumption about the quality of the policy used to collect the data, imitation learning assumes that the data has been collected by an `expert' we wish to imitate \citep{levine2020offline}.
This is useful in cases where a large amount of expert data is available, such as the case of autonomous driving \citep{pan2017agile}.
Imitation learning is clearly important during early development in biological organisms, where we learn from observing the individuals around us.
Indeed, such imitation learning is a hallmark not just of humans but has also been demonstrated in organisms as `simple' as the bumblebee \citep{loukola2017bumblebees}.
Imitation learning has also recently been used to learn models of biological neural circuits from high-resolution behavioural data \citep{aldarondo2024virtual}.

\subsection{Linear reinforcement learning}
\label{sec:linear_RL}
As we have seen in most of this tutorial, reinforcement learning is generally difficult and requires iterative algorithms that often scale poorly with the problem size.
However, there are settings where we can simplify the problem to the point where it becomes analytically tractable in an approach known as `linear reinforcement learning' \citep{todorov2006linearly, todorov2009efficient}.
This is similar to the SR approach, where we saw that the value function reduces to a linear function of the reward-per-state.
Similar to how the SR matrix can be seen as describing the dynamics of some `base policy', we also define a base policy in linear RL and compute a `control cost' as the KL divergence between transition dynamics with and without our controller:
\begin{linenomath*}
\begin{equation}
    \mathcal{L}_{ctrl}(s) = KL \left [ u(s' | s) || p(s'|s) \right ],
\end{equation}
\end{linenomath*}
where $p(s' | s)$ are the prior transition dynamics and $u(s' | s)$ are the controlled transition dynamics marginalized over the policy.
For $\mathcal{L}_{ctrl}(s)$ to be well-defined, we require $u(s'|s) = 0$ whenever $p(s'|s) = 0$, which prevents impossible transitions even under our flexible controller.
When subtracting this loss from the RL objective, the resulting utility turns out to be convex in $u$ and can therefore be solved efficiently for the controller, which implicitly specifies the policy.
This approach has recently been used as an explicit model of biological decision making \citep{piray2021linear,piray2024reconciling}.
It also has close parallels to learning and planning as inference \citep{levine2018reinforcement, solway2012goal,botvinick2012planning} and to RL with information bottlenecks \citep{lai2021policy}.
Both of these families of approaches involve reinforcement learning with a KL-regularized reward function, and they have also been used as models of biological decision making.

\subsection{Successor features}
\label{sec:SFs}
In \Cref{sec:SR}, we saw that the successor representation can be used for decision making with flexible adaptation in environments with changing reward structures.
However, we developed this framework only in the tabular setting despite extending TD-learning and Q-learning to the `deep RL' setting with function approximation.
This leaves open the question of whether a similar generalization of the SR exists.
This turns out to be the case and is known as `successor features' (SF; \citealp{barreto2017successor}), where the expected future observation of learned features of the environment are used in place of the expected future state occupancy.
Successor features have also been shown to have a biologically plausible implementation that facilitates learning and generalization in noisy and partially observable environments \citep{vertes2019neurally}.

\subsection{Multi-agent reinforcement learning}
\label{sec:multi-agent}
We have only considered the case of single agents interacting with a black-box environment.
However, in many cases, multiple agents are simultaneously interacting with each other and the environment around them \citep{gronauer2022multi}.
This means that, from the point of view of a single agent, the other agents are part of its environment.
In such settings, there are interesting learning dynamics beyond the scope of the present tutorial, but which are covered in detail by e.g. \citet{gronauer2022multi}, and which are also of substantial interest in game theory \citep{nowe2012game}.
In some cases, a whole group of agents may be working together to maximize a single joint reward function -- as is the case for members of a single sports team.
Interestingly, the learning of many individual neurons in the brain from a single common reinforcing signal (such as dopamine) can be modelled as such a multi-agent reinforcement learning problem \citep{sutton2018reinforcement}.
If the `agents' (or neurons) are assumed to have Bernoulli-logistic outputs, \citet{williams1992simple} shows that the independent learning of individual agents from the global reward signal leads to the implementation of a policy gradient algorithm at the population level \citep{sutton2018reinforcement}.

\end{document}